\documentclass[10pt]{article}
\usepackage{graphicx}

\newcommand{\beq}{\begin{equation}}
\newcommand{\eeq}{\end{equation}}
\newcommand{\beqa}{\begin{eqnarray}}
\newcommand{\eeqa}{\end{eqnarray}}
\newcommand{\non}{\nonumber}
\newcommand{\lab}{\label}
\newcommand{\bra}{\langle}
\newcommand{\ket}{\rangle}

\newcommand{\lw}[1]{\smash{\lower2.0ex\hbox{#1}}}

\begin{document}

\title{Decoherence on Grover's quantum algorithm:\\
perturbative approach}

\author{Hiroo Azuma\thanks{On leave from Canon Research Center, 5-1,
Morinosato-Wakamiya, Atsugi-shi, Kanagawa, 243-0193, Japan.}
\\
{\small Centre for Quantum Computation,
Clarendon Laboratory,}\\
{\small Parks Road, Oxford OX1 3PU, United Kingdom}\\
{\small E-mail: hiroo.azuma@qubit.org}}

\date{November 12, 2001}

\maketitle

\begin{abstract}
In this paper, we study decoherence on Grover's
quantum searching algorithm
using a perturbative method.
We assume that each two-state system (qubit)
suffers $\sigma_{z}$ error with probability $p$ ($0\leq p\leq 1$)
independently at every step in the algorithm.
Considering an $n$-qubit density operator to which Grover's operation
is applied $M$ times,
we expand it in powers of $2Mnp$ and derive its matrix element
order by order
under the
$n\rightarrow \infty$
limit.
(In this large $n$ limit,
we assume $p$ is small enough, so that $2Mnp(\geq 0)$ can take
any real positive value or $0$.)
This approach gives us an interpretation
about creation of new modes caused by $\sigma_{z}$ error
and an asymptotic form of an arbitrary order correction.
Calculating the matrix element
up to the fifth order term
numerically,
we investigate a region of $2Mnp$ (perturbative parameter)
where the algorithm finds the correct item with a threshold of
probability $P_{\mbox{\scriptsize th}}$ or more.
It satisfies
$2Mnp<(8/5)(1-P_{\mbox{\scriptsize th}})$
around $2Mnp\simeq 0$ and $P_{\mbox{\scriptsize th}}\simeq 1$,
and this linear relation is applied to
a wide range of $P_{\mbox{\scriptsize th}}$ approximately.
This observation is similar to a result
obtained by E.~Bernstein and U.~Vazirani
concerning accuracy of quantum gates
for general algorithms.
We cannot investigate a quantum to classical phase transition of
the algorithm,
because it is outside the reliable domain of our perturbation theory.
\end{abstract}

\section{Introduction}
\lab{introduction}
Since the idea of quantum computation
appeared \cite{Feynman}\cite{Deutsch}\cite{Bernstein},
a lot of researchers have been investigating its properties,
algorithms,
and implementations \cite{Deutsch-Jozsa-Simon}\cite{Cirac-Gershenfeld}.
A quantum computer can be thought a sequence of operations
which are unitary transformations and measurements
applied to two-state systems (qubits).
(The qubit means a system defined on a $2$-dimensional Hilbert space
$\{|0\ket,|1\ket\}$.)
For realizing performances that conventional (classical)
computer hardly shows,
it makes use of the properties of quantum mechanics,
such as principle of superposition and its interference,
principle of uncertainty,
and entanglement
(quantum correlation which is stronger than classical one).

One of the most serious problem for realizing
quantum computation is decoherence,
which is caused by an interaction
between the system of quantum computer
and an environment that surrounds it \cite{Zurek}\cite{Unruh}.
It is pointed out that
quantum information stored as a quantum state is
fragile and collapses at ease by this disturbance.
To investigate it,
some decoherence processes are assumed
and their effects on quantum algorithms are
estimated \cite{Chuang-Laflamme}\cite{Barenco}.
For overcoming these troubles,
quantum error-correcting codes are proposed
and their availability is examined \cite{Shor-Steane-Calderbank}\cite{Mussinger}.

Not only for practical purposes
but also for theoretical interests,
it is an important question
how robust the quantum algorithm is against this disturbance.
We can expect that the quantum computer loses its efficiency gradually
as decoherence gets stronger.
Some researchers regard it as
a quantum to classical phase transition \cite{Aharonov}.

Grover's algorithm is considered to be
an efficient amplitude amplification process
for quantum states,
so that it is often called a searching
algorithm \cite{Grover}\cite{Boyer-Brassard}.
By applying the same unitary transformation to the state in iteration and
amplifying an amplitude of one basis vector that we want gradually,
Grover's algorithm picks up it from a uniform superposition of
$2^{n}$ basis vectors with certain probability by $O(2^{n/2})$ steps.
Because it handles a general problem
(an unsorted database search),
it can be formulated as an oracle problem,
and it is proved
that its efficiency is optimal in view of computational time
(the number of queries for the oracle) \cite{Boyer-Brassard}\cite{Ambainis},
many researchers have analysed this algorithm precisely and proposed
a lot of applications \cite{Brassard-Carlini}.

In this paper,
we study the decoherence on Grover's quantum algorithm
with a perturbative method.
We assume a simple model
and investigate it for higher order perturbation
(numerically up to the fifth order correction),
under the limit of an infinite number of qubits.

The model has the following three characteristics.
First, in the Grover's algorithm,
we assume that we search the basis vector of $|0\cdots 0\ket$
from the uniform superposition of the $n$-qubit logical basis.
This assumption simplifies the iterated transformation.
Second, we assume each qubit interacts with the environment independently
and suffers a phase damping
which causes $\sigma_{z}$ error with probability $p$
and does nothing with probability $(1-p)$ \cite{Huelga}.
Third, we take the limit of $n\rightarrow \infty$,
so that the matrix element of the density operator is simplified.

In our perturbation theory,
we expand an $n$-qubit density operator
to which Grover's operation is applied $M$ times
in powers of $2Mnp$ \cite{Schiff-Ramond-Halzen}.
Investigating higher order terms of the perturbation,
we obtain a physical interpretation that the $\sigma_{z}$ error
creates new modes as the algorithm goes steps.
When we take the large $n$ limit mentioned above,
we assume $p$ is small enough,
so that a perturbative parameter $2Mnp$ can take any positive value or $0$.
Taking the $n\rightarrow \infty$ limit simplifies
the matrix element of the density operator
and gives us an asymptotic form of an arbitrary order term.

Calculating the matrix element
up to the fifth order term
numerically,
we investigate a range of $2Mnp$
where the algorithm finds the correct item with a threshold of
probability $P_{\mbox{\scriptsize th}}$ or more.
It satisfies
$2Mnp<(8/5)(1-P_{\mbox{\scriptsize th}})$
around $2Mnp\simeq 0$ and $P_{\mbox{\scriptsize th}}\simeq 1$,
and this linear relation can be approximately
applied to a wide range of
$P_{\mbox{\scriptsize th}}$.
Hence, if we fix $P_{\mbox{\scriptsize th}}$ to a certain value
($P_{\mbox{\scriptsize th}}=1/2$,
for example),
we have to suppress the error rate to
a value which is in proportion to
the inverse of
the number of quantum gates.

Similar results are obtained by E.~Bernstein and U.~Vazirani
in the study of accuracy for quantum gates \cite{Bernstein}.
They consider a quantum circuit where each quantum gate has
a constant error because of inaccuracy,
so it is an error of a unitary transformation and
it never causes dissipation to the quantum computer.
They estimate inaccuracy $\epsilon$
for which the quantum algorithm is available
against the fixed number of time steps $T$,
and obtain $2T\epsilon< 1-P_{\mbox{\scriptsize th}}$.
If we regard $p/2$ as inaccuracy $\epsilon$,
and $2Mn$ as the number of whole steps in algorithm $T$,
it is similar to our observations
except for a factor.

A.~Barenco et al. study the approximate
quantum Fourier transformation (AQFT)
and its decoherence \cite{Barenco}.
Although motivation is slightly different from
E.~Bernstein and U.~Vazirani's,
we can think their model to be the quantum Fourier transformation (QFT)
with inaccurate phase gates.
They confirmed that AQFT can make a performance
that is not so worse than QFT's one.

This article is arranged as follows.
In Section~\ref{Section-model-decoherence},
we describe the model that we analyse in this paper.
In Section~\ref{section-perturbative-method},
we formulate a perturbation theory for our model
and explain physical quantities that we estimate.
In Section~\ref{Section-Matrix-elements-density-operator-0th-first},
we derive the matrix element of the density operator
of the quantum computer for the $0$-th and first order.
We give a physical interpretation about creation of
new modes by $\sigma_{z}$ errors
in Section~\ref{Section-physical-interpretation-multi-particle-creation}.
Then, we derive the second order correction of the matrix element
in Section~\ref{Section-Matrix-elements-density-operator-second}.
In Section~\ref{Section-large-n-qubits-limit},
we take the limit of an infinite number of qubits
and give the asymptotic form of an arbitrary
order term.
In Section~\ref{Section-numerical-calculations},
we carry out numerical calculations of physical quantities
up to the fifth order correction.
In Section~\ref{Section-discussions},
we give brief discussions concerned with our results.
We collect formulas for deriving matrix elements
in Appendix~\ref{Appendix-formulas-for-calculating-matrix-elements},
and give some notes about numerical calculations
of higher order perturbative terms
in Appendix~\ref{Appendix-notes-for-numerical-calculations}.

\section{Model of decoherence}
\lab{Section-model-decoherence}
In this section, we describe a model that we analyse.
It is a quantum process of Grover's algorithm
which suffers a phase error in iteration.

At first, we give a brief review of Grover's algorithm \cite{Grover}.
Starting from the $n$-qubit ($n\geq 2$) uniform superposition
on a logical basis,
\beq
W|0\cdots 0\ket
=
\frac{1}{\sqrt{2^{n}}}
\sum_{x\in\{0,1\}^{n}}|x\ket,
\lab{n-qubit-uniform superposition}
\eeq
it increases gradually an amplitude of a certain basis vector
$|x_{0}\ket$ ($x_{0}\in\{0,1\}^{n}$)
which is indicated by a quantum oracle.
An operator $W$ in Eq.~(\ref{n-qubit-uniform superposition})
is an $n$-fold product of a one-qubit unitary transformation
(Hadamard transformation)
and given by
$W=H^{\otimes n}$,
where
\beqa
H=H^{\dagger}
=\frac{1}{\sqrt{2}}
\bordermatrix{
       & \bra 0| & \bra 1| \cr
|0\ket & 1       & 1       \cr
|1\ket & 1       & -1      \cr
},
\quad\quad
H^{\dagger}H=\mbox{\boldmath $I$}.
\eeqa
The quantum oracle can be regarded as a black box,
and actually it is a quantum gate
which shifts phases of logical basis vectors as
\beq
R_{x_{0}}:
\quad
\left\{
\begin{array}{llll}
|x_{0}\ket & \rightarrow & -|x_{0}\ket & \\
|x\ket & \rightarrow & |x\ket & \mbox{for $x\neq x_{0}$}
\end{array}
\right. ,
\lab{selective-phase-shift-operator}
\eeq
where $x_{0},x \in\{0,1\}^{n}$,
$R_{x_{0}}^{\dagger}=R_{x_{0}}$,
and $R_{x_{0}}^{\dagger}R_{x_{0}}=\mbox{\boldmath $I$}$.

To let probability of observing $|x_{0}\ket$
be greater than a certain value ($1/2$, for example),
we repeat the following procedure $O(\sqrt{2^{n}})$ times.
\begin{enumerate}
\item Apply $R_{x_{0}}$ to the $n$-qubit state.
\item Apply $D=WR_{0}W$ to the $n$-qubit state.
\end{enumerate}
$R_{0}$ is a selective phase shift operator
which multiplies a factor $(-1)$ to $|0\cdots 0\ket$
and does nothing to the other basis vectors,
as defined in Eq.~(\ref{selective-phase-shift-operator}).
$D$ is called the inversion about average operation.

From now on,
we assume that
we amplify an amplitude of $|0\cdots 0\ket$.
From this assumption,
we can write an operation iterated in the algorithm as
\beq
DR_{0}=(WR_{0}W)R_{0}.
\eeq
After repeating this operation $M$ times
from the initial state of $W|0\ket$($=W|0\cdots 0\ket$),
we obtain the state of $(WR_{0})^{2M}W|0\ket$.
(We often write $|0\ket$
as an abbreviation of the $n$-qubit state
$|0\cdots 0\ket$ for a simple notation.)

Next, we think about the decoherence.
In this paper, we consider the following
one-qubit phase error \cite{Huelga},
\beq
\rho
\rightarrow
\rho'=p\sigma_{z}\rho\sigma_{z}+(1-p)\rho
\quad\quad
\mbox{for $0\leq p \leq 1$},
\lab{one-qubit-phase-error}
\eeq
where $\rho$ is an arbitrary one-qubit density operator
and $\sigma_{z}$ is one of the Pauli matrices given by
\beqa
\sigma_{z}=\sigma_{z}^{\dagger}
=
\bordermatrix{
       & \bra 0| & \bra 1| \cr
|0\ket & 1       & 0       \cr
|1\ket & 0       & -1      \cr
}.
\eeqa
For simplicity,
we assume that
the phase error of Eq.~(\ref{one-qubit-phase-error})
occurs in each qubit of the state independently
before every $R_{0}$ operation during the algorithm.
It assumes that
each qubit interacts with its environment
and suffers the phase error independently.

Here,
we add some notes.
First, because $R_{0}\in U(2^{n})$ is
applied to all $n$ qubits
and
$H\in U(2)$ is applied to only one qubit,
we can imagine that
the realization of $R_{0}$
is more difficult than that of $W=H^{\otimes n}$.
Hence, we assume that the phase error occurs only before $R_{0}$.
Second,
although we assume a very simple error defined
in Eq.~(\ref{one-qubit-phase-error}),
we can think other complicated errors.
For example, we can consider a phase error
caused by an interaction
between the environment and two qubits,
and it may occur with probability of $O(p^{2})$.
In this paper,
we do not assume such complicated errors.
We discuss how the disturbance of Eq.~(\ref{one-qubit-phase-error})
occurs in Section~\ref{Subsection-how-the-phase-error-occurs}.

\section{The perturbative method}
\lab{section-perturbative-method}
Let $\rho^{(M)}$ be the density matrix which is obtained
by applying Grover's operation $M$ times
to the $n$-qubit initial state $W|0\ket$.
The decoherence of Eq.~(\ref{one-qubit-phase-error})
occurs $2Mn$ times in $\rho^{(M)}$.
We can expand $\rho^{(M)}$ in powers
of $p$ and $(1-p)$ as follows,
\beqa
\rho^{(M)}
&=&
(1-p)^{2Mn}T_{0}^{(M)}
+(1-p)^{2Mn-1}p T_{1}^{(M)}
+\cdots \non \\
&=&
\sum_{h=0}^{2Mn}(1-p)^{2Mn-h}p^{h} T_{h}^{(M)},
\lab{definition-(1-p)-p-expanded-density-operator}
\eeqa
where $\{T_{h}^{(M)}\}$ are given by
\beqa
T_{0}^{(M)}
&=&
(WR_{0})^{2M}W|0\ket\bra 0|W(R_{0}W)^{2M},
\lab{definition-T0M} \\
T_{1}^{(M)}
&=&
\sum_{i=1}^{n}
\sum_{k=0}^{2M-1}
(WR_{0})^{2M-k}\sigma_{z}^{(i)}(WR_{0})^{k}W|0\ket
\bra 0|W(R_{0}W)^{k}\sigma_{z}^{(i)}(R_{0}W)^{2M-k},
\lab{definition-T1M} \\
T_{2}^{(M)}
&=&
\sum_{i=1}^{n}
\sum_{{j=1}\atop{i<j}}^{n}
\sum_{k=0}^{2M-1}
(WR_{0})^{2M-k}\sigma_{z}^{(i)}\sigma_{z}^{(j)}
(WR_{0})^{k}W|0\ket\bra \mbox{h.c.}| \non \\
&&
+\sum_{i=1}^{n}
\sum_{j=1}^{n}
\sum_{k=0}^{2M-1}
\sum_{l=1}^{2M-k-1}
(WR_{0})^{2M-k-l}\sigma_{z}^{(i)}(WR_{0})^{l}\sigma_{z}^{(j)}
(WR_{0})^{k}W|0\ket\bra \mbox{h.c.}|,
\lab{definition-T2M}
\eeqa
and so on,
$\sigma_{z}^{(i)}$ represents the operator applied to the $i$-th
qubit ($1\leq i\leq n$),
and $\bra \mbox{h.c.}|$ represents
a hermitian conjugation of the ket vector
on its left side.
We can regard $T_{h}^{(M)}$ as a density operator
whose trace is not normalised.
It represents states
where $h$ errors occur
during the iteration of $M$ operations.

On the other hand, we expand $\rho^{(M)}$
in powers of $p$ as follows,
\beqa
\rho^{(M)}
&=&
\rho_{0}^{(M)}
+2Mnp\rho_{1}^{(M)}
+\frac{1}{2}(2Mn)(2Mn-1)p^{2}\rho_{2}^{(M)}
+\cdots \non \\
&=&
\sum_{h=0}^{2Mn}
{2Mn\choose{h}}
p^{h}\rho_{h}^{(M)},
\lab{density-operator-expansion-p-powers}
\eeqa
where
\beqa
\rho_{0}^{(M)}
&=&
T_{0}^{(M)}, \non \\
\rho_{1}^{(M)}
&=&
-T_{0}^{(M)}+\frac{T_{1}^{(M)}}{2Mn}, \non \\
\rho_{h}^{(M)}
&=&
(-1)^{h}\sum_{j=0}^{h}(-1)^{j}
{2Mn\choose{j}}^{-1}
{h\choose{j}}
T_{j}^{(M)}
\quad\quad
\mbox{for $h=0,1,\cdots,2Mn$}.
\lab{Definitions-p-expanded-density-operators-0123}
\eeqa
In Section~\ref{Section-large-n-qubits-limit},
we take a large $n$ limit (the limit of an infinite number of qubits).
If $2Mnp$ is small enough,
we can consider the series of Eq.~(\ref{density-operator-expansion-p-powers}) to be a
perturbative expansion.
Because we divide $T_{h}^{(M)}$ by $(Mn)^{h}$
as Eq.~(\ref{Definitions-p-expanded-density-operators-0123}),
an expectation value
of $\rho_{h}^{(M)}$ ($h=0,1,\cdots,2Mn$)
can converge on a finite value in the limit of $n\rightarrow \infty$.

\begin{figure}
\begin{center}
\includegraphics[scale=1.0]{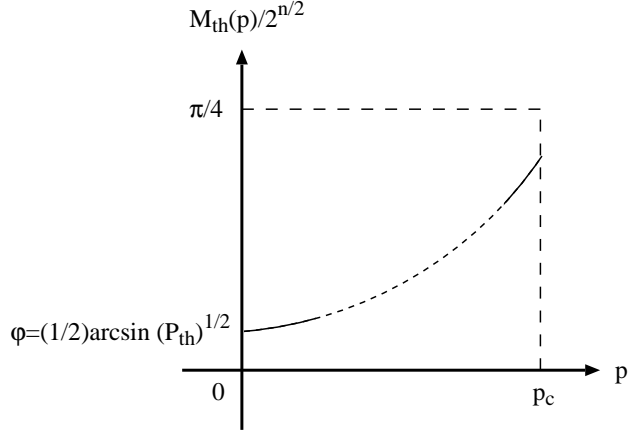}
\end{center}
\caption{Variation of $M_{\mbox{\scriptsize th}}(p)/\sqrt{2^{n}}$ against $p$
with the threshold probability $P_{\mbox{\scriptsize th}}$
under large but finite $n$.}
\lab{critical-point}
\end{figure}

With these preparations,
we will investigate the following physical quantities.
Let $P_{\mbox{\scriptsize th}}$ be a threshold of probability
($0< P_{\mbox{\scriptsize th}}\leq 1$),
so that if the quantum computer finds the item that we want
(in our model, it is $|0\ket$) with the probability $P_{\mbox{\scriptsize th}}$ or more,
we regard it available,
and otherwise we do not consider it available.
Then, we consider the least number of the operations
that we need to repeat for amplifying the probability of
observing $|0\ket$ to $P_{\mbox{\scriptsize th}}$ or more
for given $p$.
We can describe it as
$M=M_{\mbox{\scriptsize th}}(p;P_{\mbox{\scriptsize th}})$,
and it satisfies
$\bra 0|\rho^{(M)}|0\ket=P_{\mbox{\scriptsize th}}$.
(For convenience,
we write it as
$M_{\mbox{\scriptsize th}}(p)$
with omitting $P_{\mbox{\scriptsize th}}$
as far as it does not make confusion.)
Because of
\beq
\bra 0|T_{0}^{(M)}|0\ket_{p=0}
\simeq
\sin^{2}(2M/\sqrt{2^{n}})
\quad\quad
\mbox{for large but finite $n$},
\lab{T0M-large-n-limit}
\eeq
obtained
in Section~\ref{Section-Matrix-elements-density-operator-0th-first},
$M_{\mbox{\scriptsize th}}(p)$ takes a value of $\varphi\sqrt{2^{n}}$
at $p=0$ (with no decoherence)
for large finite $n$,
where $\varphi=(1/2)\arcsin \sqrt{P_{\mbox{\scriptsize th}}}$.

As $p$ gets larger,
we can expect that $M_{\mbox{\scriptsize th}}(p)$ increases monotonously.
It could be possible
for certain $p_{\mbox{\scriptsize c}}$ or more that
we never observe $|0\ket$ at least
with probability of $P_{\mbox{\scriptsize th}}$.
(Hence, $p_{\mbox{\scriptsize c}}$ depends on
$P_{\mbox{\scriptsize th}}$.)
Such behaviour of $M_{\mbox{\scriptsize th}}(p)$ can be drawn in
Figure~\ref{critical-point}.
We multiply a factor $1/\sqrt{2^{n}}$ to $M_{\mbox{\scriptsize th}}(p)$ for normalisation.
Because $\bra 0|\rho^{(M)}|0\ket$
with $p=0$ increases monotonously
from $M=0$ to $M=(\pi/4)\sqrt{2^{n}}$ 
and then decreases as shown in Eq.~(\ref{T0M-large-n-limit}),
$M_{\mbox{\scriptsize th}}(p)/\sqrt{2^{n}}$
at $p=p_{\mbox{\scriptsize c}}$ is
equal to or less than $\pi/4$.

\begin{figure}
\begin{center}
\includegraphics[scale=1.0]{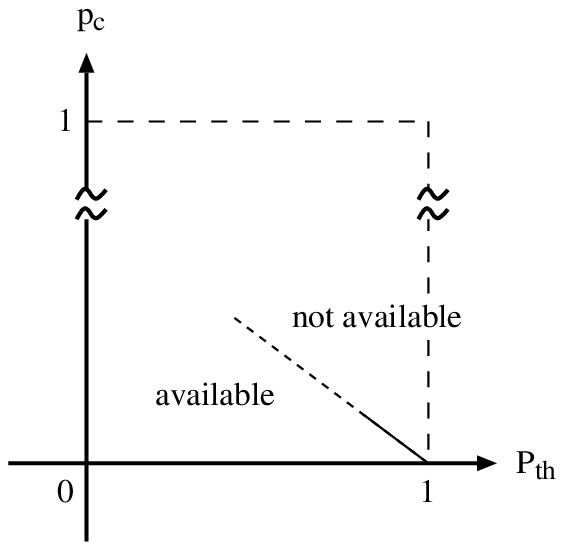}
\end{center}
\caption{Variation of $p_{\mbox{\scriptsize c}}$
against $P_{\mbox{\scriptsize th}}$.}
\lab{qc-available}
\end{figure}

Regarding $P_{\mbox{\scriptsize th}}$ as a threshold
whether the quantum computer is available or not,
we can consider $p_{\mbox{\scriptsize c}}$ to be a critical point.
(This is not a so-called quantum to classical phase transition.)
We can draw a graph of $p_{\mbox{\scriptsize c}}$ against $P_{\mbox{\scriptsize th}}$
as Figure~\ref{qc-available}.
(If $P_{\mbox{\scriptsize th}}=1$, we obtain $p_{\mbox{\scriptsize c}}=0$.)

In this paper, we calculate physical quantities
using the perturbative parameter $x=2Mnp$,
so that we take $M/\sqrt{2^{n}}$ and $x$ for independent variables.
(In our original model defined in Section~\ref{Section-model-decoherence},
we take $M$ and $p$ for independent variables.)
We can define as well
$\tilde{M}_{\mbox{\scriptsize th}}(x; P_{\mbox{\scriptsize th}})/\sqrt{2^{n}}$
that represents the least number of the operations iterated
for amplifying the probability of $|0\ket$ to $P_{\mbox{\scriptsize th}}$
for given $x$.
Furthermore, we also obtain $x_{c}$,
for which or more we can never detect $|0\ket$ at least
with probability $P_{\mbox{\scriptsize th}}$.
Hence, we obtain a graph of
$\tilde{M}_{\mbox{\scriptsize th}}(x)/\sqrt{2^{n}}$
versus $x$
instead of Figure~\ref{critical-point},
and
that of 
$x_{\mbox{\scriptsize c}}$
versus $P_{\mbox{\scriptsize th}}$
instead of Figure~\ref{qc-available}.
The differences of these quantities are discussed
in Section~\ref{Section-numerical-calculations}.

The dependence of $x_{\mbox{\scriptsize c}}$
on $P_{\mbox{\scriptsize th}}$
gives us useful information.
If we regard $2Mn$
as the number of computational steps $T$,
and $p/2$ as a parameter $\epsilon$ which represents a degree of errors,
it serves the region of
$T\epsilon$
where the quantum computer is available for $P_{\mbox{\scriptsize th}}$,
because of $x=2Mnp$.
To make these analyses,
we need to know $\bra 0|\rho^{(M)}|0\ket$.
In the following sections, we calculate $\bra 0|\rho^{(M)}|0\ket$
to the fifth order of $p$
(up to $(2Mnp)^{5}$)
numerically
under the $n\rightarrow \infty$ limit.

\section{Matrix elements up to the first order}
\lab{Section-Matrix-elements-density-operator-0th-first}
In this section,
we consider matrix elements of the density operators
with no and one error,
$\bra 0|T_{0}^{(M)}|0\ket$ and $\bra 0|T_{1}^{(M)}|0\ket$,
defined in Eqs.~(\ref{definition-T0M}) and (\ref{definition-T1M}).

First, we derive $T_{0}^{(M)}$ and $\bra 0|T_{0}^{(M)}|0\ket$.
Using Eq.~(\ref{Appendix-WR0-2k-W0})
in Appendix~\ref{Appendix-subsec-formulas-of-WR0-k-W-0},
we obtain
\beq
T_{0}^{(M)}
=
[\sin(2M+1)\theta|0\ket
+
\frac{\cos(2M+1)\theta}{\sqrt{2^{n}-1}}
\sum_{x\neq 0}|x\ket]
\bra \mbox{h.c.}|,
\lab{density-operator-T0}
\eeq
\beq
\bra 0|T_{0}^{(M)}|0\ket
=\sin^{2}(2M+1)\theta,
\lab{matrix-element-T0}
\eeq
where
\beq
\sin\theta=\frac{1}{\sqrt{2^{n}}},
\quad\quad
\cos\theta=\sqrt{\frac{2^{n}-1}{2^{n}}}.
\lab{Boyer-parameter-theta}
\eeq
(This parameter $\theta$ is introduced by M.~Boyer et al.
and it simplifies our notation \cite{Boyer-Brassard}.)
From Eq.~(\ref{matrix-element-T0}),
we notice the following facts.
If there is no decoherence ($p=0$),
we can amplify probability of
observing $|0\ket$ to unity.
Taking large (but finite) $n$,
we obtain $\sin\theta\simeq\theta$
and $\theta\simeq 1/\sqrt{2^{n}}$,
and
we can observe $|0\ket$ with probability of unity
after repeating Grover's operation
$M_{\mbox{\scriptsize max}}\simeq (\pi/4)\sqrt{2^{n}}$ times.

Then, we think about $T_{1}^{(M)}$ and $\bra 0|T_{1}^{(M)}|0\ket$.
For convenience,
in spite of the definition of $T_{1}^{(M)}$ given in
Eq.~(\ref{definition-T1M}),
we rewrite it as follows,
\beqa
T_{1}^{(M)}
&=&
\sum_{i=1}^{n}
\sum_{k=0}^{M-1}
(WR_{0})^{2(M-k)}\sigma_{z}^{(i)}(WR_{0})^{2k}W|0\ket
\bra \mbox{h.c.}| \non \\
&&
+\sum_{i=1}^{n}
\sum_{k=0}^{M-1}
(WR_{0})^{2(M-k)-1}\sigma_{z}^{(i)}(WR_{0})^{2k+1}W|0\ket
\bra \mbox{h.c.}|.
\lab{definition-T1M-odd-even}
\eeqa
We can derive an explicit form of $\bra 0|T_{1}^{(M)}|0\ket$
using formulas collected
in Appendixes~\ref{Appendix-subsec-formulas-of-WR0-k-W-0}
and \ref{Appendix-subsec-formulas-of-R0W-k-0}.
The matrix element of the first term
in Eq.~(\ref{definition-T1M-odd-even})
can be given by
$|{\cal T}_{\mbox{\scriptsize even}}^{(k)}|^{2}$, where
\beqa
{\cal T}_{\mbox{\scriptsize even}}^{(k)}
&\equiv&
\bra 0|(WR_{0})^{2(M-k)}\sigma_{z}^{(i)}(WR_{0})^{2k}W|0\ket \non \\
&=&
(-1)^{M-k}
[\cos 2(M-k)\theta \bra 0|
+
\frac{\sin 2(M-k)\theta}{\sqrt{2^{n}-1}}\sum_{x\neq 0}\bra x|]
\sigma_{z}^{(i)} \non \\
&&
\times
(-1)^{k}
[\sin(2k+1)\theta|0\ket
+
\frac{\cos(2k+1)\theta}{\sqrt{2^{n}-1}}
\sum_{y\neq 0}|y\ket] \non \\
&=&
(-1)^{M}
[\cos 2(M-k)\theta \sin(2k+1)\theta
-\frac{1}{2^{n}-1}\sin 2(M-k)\theta\cos(2k+1)\theta].
\lab{Matrix-element-T1-even}
\eeqa
We notice that ${\cal T}_{\mbox{\scriptsize even}}^{(k)}$
does not depend on the subscript $i$.
In a similar way, we can obtain
a matrix element of the second term
in Eq.~(\ref{definition-T1M-odd-even})
as
$|{\cal T}_{\mbox{\scriptsize odd}}^{(k)}|^{2}$, where
\beqa
{\cal T}_{\mbox{\scriptsize odd}}^{(k)}
&\equiv&
\bra 0|(WR_{0})^{2(M-k)-1}\sigma_{z}^{(i)}(WR_{0})^{2k+1}W|0\ket \non \\
&=&
(-1)^{M}
[\sin (2M-2k-1)\theta \cos 2(k+1)\theta
-\frac{1}{2^{n}-1}\cos (2M-2k-1)\theta\sin 2(k+1)\theta].
\lab{Matrix-element-T1-odd}
\eeqa
(Here, we notice
$|{\cal T}_{\mbox{\scriptsize even}}^{(k)}|^{2}
=
|{\cal T}_{\mbox{\scriptsize odd}}^{(M-k-1)}|^{2}$.)
Hence, we obtain
\beq
\bra 0|T_{1}^{(M)}|0\ket
=
n\sum_{k=0}^{M-1}
(|{\cal T}_{\mbox{\scriptsize even}}^{(k)}|^{2}
+
(|{\cal T}_{\mbox{\scriptsize odd}}^{(k)}|^{2}).
\lab{Explicit-form-matrix-element-T1M}
\eeq

\section{Physical interpretation for creation of modes}
\lab{Section-physical-interpretation-multi-particle-creation}
If we derive the explicit form of $T_{1}^{(M)}$,
we can obtain a physical interpretation of creating modes
by $\sigma_{z}$ error.

Let us consider the state included
in Eq.~(\ref{definition-T1M-odd-even}),
\beq
(WR_{0})^{l}\sigma_{z}^{(i)}(WR_{0})^{k}W|0\ket,
\lab{WR0-l-1-phase-error-WR0-k-W0}
\eeq
which suffers the phase error
only once.
From Eq.~(\ref{Appendix-WR0-2k-W0})
in Appendix~\ref{Appendix-subsec-formulas-of-WR0-k-W-0},
we obtain the following equation,
\beqa
\lefteqn{\sigma_{z}^{(i)}(WR_{0})^{2k}W|0\ket} \non \\
&=&
(-1)^{k}
[\sin(2k+1)\theta|0\ket
+
\frac{\cos(2k+1)\theta}{\sqrt{2^{n}-1}}
(\sum_{{x\neq 0}\atop{x_{i}=0}}|x\ket
-\sum_{{x\neq 0}\atop{x_{i}=1}}|x\ket)] \non \\
&=&
(WR_{0})^{2k}W|0\ket
-\sqrt{2}(-1)^{k}
\frac{\cos(2k+1)\theta}{\cos\theta}|\eta_{i}\ket
\quad\quad
\mbox{for $k=0,1,\cdots$},
\lab{1-phase-error-WR0-2k-W0}
\eeqa
where
\beq
|\eta_{i}\ket
=\frac{1}{\sqrt{2^{n-1}}}
\sum_{x_{i}=1}|x\ket
\quad\quad
\mbox{for $i=1,\cdots,n$}.
\lab{definition-eta-i}
\eeq
In a similar way, from Eq.~(\ref{Appendix-WR0-(2k+1)-W0}),
we obtain
\beq
\sigma_{z}^{(i)}(WR_{0})^{2k+1}W|0\ket
=
(WR_{0})^{2k+1}W|0\ket
+\sqrt{2}(-1)^{k}
\frac{\sin 2(k+1)\theta}{\cos\theta}|\eta_{i}\ket
\quad\quad
\mbox{for $k=0,1,\cdots$}.
\lab{1-phase-error-WR0-(2k+1)-W0}
\eeq

To obtain an explicit form of
Eq.~(\ref{WR0-l-1-phase-error-WR0-k-W0}),
we have to apply $WR_{0}$ to
Eqs.~(\ref{1-phase-error-WR0-2k-W0})
and (\ref{1-phase-error-WR0-(2k+1)-W0})
from the left side step by step.
(We count $WR_{0}$ as one step for a while.)
We can derive an explicit form of $(WR_{0})^{l}|\eta_{i}\ket$
as follows.
In the case of $l=1$, we obtain
\beq
WR_{0}|\eta_{i}\ket
=
\frac{1}{\sqrt{2}}
(|0\ket-|\overline{i}\ket),
\eeq
where $\overline{i}$ represents a binary string
whose all digits are `$0$' but the $i$-th digit is `$1$',
so that
\beq
|\overline{i}\ket
=|0
\cdots 0
\begin{array}[t]{c}
1 \\
\uparrow \\
i
\end{array}
0
\cdots 0\ket.
\lab{definition-overline-i}
\eeq
For $l=2$, we obtain
\beqa
(WR_{0})^{2}|\eta_{i}\ket
&=&
\frac{1}{\sqrt{2}}
WR_{0}
(|0\ket-|\overline{i}\ket) \non \\
&=&
-\frac{1}{\sqrt{2^{n-1}}}
\sum_{x_{i}=0}|x\ket \non \\
&=&
-\sqrt{2}W|0\ket+|\eta_{i}\ket.
\eeqa
Therefore, we obtain
\beqa
(WR_{0})^{2l}|\eta_{i}\ket
&=&
-\sqrt{2}\sum_{m=0}^{l-1}(WR_{0})^{2m}W|0\ket+|\eta_{i}\ket
\quad\quad
\mbox{for $l=0,1,\cdots$},
\lab{WR0-2k-eta-i} \\
(WR_{0})^{2l+1}|\eta_{i}\ket
&=&
-\sqrt{2}\sum_{m=0}^{l-1}(WR_{0})^{2m+1}W|0\ket
+\frac{1}{\sqrt{2}}
(|0\ket-|\overline{i}\ket)
\quad\quad
\mbox{for $l=0,1,\cdots$},
\lab{WR0-2k+1-eta-i}
\eeqa
where $\sum_{l=0}^{-1}$ means that no term is summed up.

From
Eqs.~(\ref{1-phase-error-WR0-2k-W0}),
(\ref{1-phase-error-WR0-(2k+1)-W0}),
(\ref{WR0-2k-eta-i}),
and (\ref{WR0-2k+1-eta-i}),
we can describe
$(WR_{0})^{l}\sigma_{z}^{(i)}(WR_{0})^{k}W|0\ket$
for $k=0,1,\cdots$ and $l=1,2,\cdots$
with $(WR_{0})^{k}W|0\ket$, $|\eta_{i}\ket$,
and $(|0\ket-|\overline{i}\ket)$.
(We obtain the explicit form of
$(WR_{0})^{k}W|0\ket$
in Appendix~\ref{Appendix-subsec-formulas-of-WR0-k-W-0}.)
For example,
we can write
\beqa
\lefteqn{(WR_{0})^{2l}\sigma_{z}^{(i)}(WR_{0})^{2k}W|0\ket} \non \\
&=&
(WR_{0})^{2(k+l)}W|0\ket
-\sqrt{2}(-1)^{k}
\frac{\cos(2k+1)\theta}{\cos\theta}
[-\sqrt{2}\sum_{m=0}^{l-1}(WR_{0})^{2m}W|0\ket+|\eta_{i}\ket] \non \\
&&\quad\quad
\mbox{for $k=0,1,\cdots$ and $l=1,2,\cdots$}.
\lab{WR0-2l-sigmaZ-WR0-2k-W0}
\eeqa

\begin{figure}
\begin{center}
\includegraphics[scale=1.0]{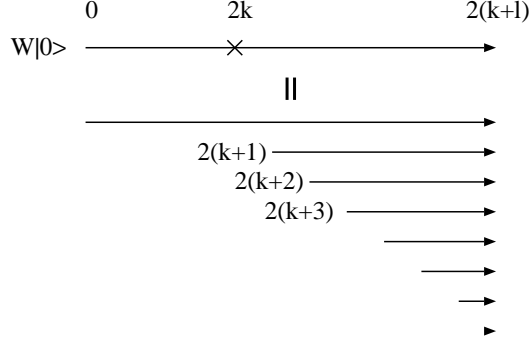}
\end{center}
\caption{Creation of modes by the error.}
\lab{multi-creation}
\end{figure}

This equation allows us the following interpretation.
If $\sigma_{z}$ error occurs in the $i$-th qubit
of the $n$-qubit state at the $2k$-th step,
it causes new modes which are created as the initial
state $W|0\ket$
at every two steps from $2(k+1)$, that is,
$2(k+1)$-th, $2(k+2)$-th, $2(k+3)$-th, $\cdots$,
and so on.
(See Figure~\ref{multi-creation}.)
The state becomes a superposition of them.

Here, we derive the matrix element
$\bra 0|T_{1}^{(M)}|0\ket$ again
using this interpretation.
From Eqs.~(\ref{WR0-2l-sigmaZ-WR0-2k-W0})
and (\ref{Appendix-WR0-2k-W0})
in Appendix~\ref{Appendix-subsec-formulas-of-WR0-k-W-0},
we obtain
\beqa
\lefteqn{\bra 0|(WR_{0})^{2(M-k)}
\sigma_{z}^{(i)}(WR_{0})^{2k}W|0\ket} \non \\
&=&
\bra 0|(WR_{0})^{2M}W|0\ket \non \\
&&
-\sqrt{2}(-1)^{k}
\frac{\cos(2k+1)\theta}{\cos\theta}
[-\sqrt{2}\sum_{l=0}^{M-k-1}\bra 0|(WR_{0})^{2l}W|0\ket
+\bra 0|\eta_{i}\ket] \non \\
&=&
(-1)^{M}\sin(2M+1)\theta
+2(-1)^{k}
\frac{\cos(2k+1)\theta}{\cos\theta}
\sum_{l=0}^{M-k-1}(-1)^{l}\sin(2l+1)\theta.
\lab{Expansion-0-WR0-2(M-k)-sigmaZ-WR0-2k-W0}
\eeqa
(We eliminate one $\sigma_{z}$ operator
from the above equation.)
Then using a formula of Eq.~(\ref{Appendix-sum-sin-2l+1-theta})
in
Appendix~\ref{Appendix-subsec-formulas-for-summation-trigonometric-functions}
to sum up trigonometric functions,
we can rewrite Eq.~(\ref{Expansion-0-WR0-2(M-k)-sigmaZ-WR0-2k-W0}) as
\beqa
\lefteqn{\bra 0|(WR_{0})^{2(M-k)}
\sigma_{z}^{(i)}(WR_{0})^{2k}W|0\ket} \non \\
&=&
(-1)^{M}\sin(2M+1)\theta
+2(-1)^{k}
\frac{\cos(2k+1)\theta}{\cos\theta}
\frac{(-1)^{M-k-1}}{2}
\frac{\sin 2(M-k)\theta}{\cos\theta} \non \\
&=&
(-1)^{M}
[\cos 2(M-k)\theta\sin(2k+1)\theta
-\frac{1}{2^{n}-1}
\sin 2(M-k)\theta\cos(2k+1)\theta] \non \\
&=&
{\cal T}_{\mbox{\scriptsize even}}^{(k)},
\eeqa
where we substitute
$\cos^{2}\theta=(2^{n}-1)/2^{n}$ of
Eq.~(\ref{Boyer-parameter-theta}).
In this derivation,
we expand the matrix elements into a series of modes
by Eq.~(\ref{WR0-2l-sigmaZ-WR0-2k-W0}),
and sum up them by the formula of
Appendix~\ref{Appendix-subsec-formulas-for-summation-trigonometric-functions}.
We often use this technique in this paper.

In a similar way,
using Eqs.~(\ref{1-phase-error-WR0-(2k+1)-W0}),
(\ref{WR0-2k+1-eta-i}),
(\ref{Appendix-WR0-(2k+1)-W0}),
and (\ref{Appendix-sum-cos-2(l+1)-theta}),
we obtain
\beqa
\lefteqn{\bra 0|(WR_{0})^{2(M-k)-1}
\sigma_{z}^{(i)}(WR_{0})^{2k+1}W|0\ket} \non \\
&=&
\bra 0|(WR_{0})^{2M}W|0\ket \non \\
&&
+\sqrt{2}(-1)^{k}
\frac{\sin 2(k+1)\theta}{\cos\theta}
[-\sqrt{2}\sum_{l=0}^{M-k-2}\bra 0|(WR_{0})^{2l+1}W|0\ket+
\frac{1}{\sqrt{2}}\bra 0|(|0\ket-|\overline{i}\ket)] \non \\
&=&
(-1)^{M}\sin(2M+1)\theta
+2(-1)^{k}
\frac{\sin 2(k+1)\theta}{\cos\theta}
[-\sum_{l=0}^{M-k-2}(-1)^{l}\cos 2(l+1)\theta+\frac{1}{2}] \non \\
&=&
(-1)^{M}\sin(2M+1)\theta \non \\
&&
+2(-1)^{k}
\frac{\sin 2(k+1)\theta}{\cos\theta}
[-\frac{1}{2}-
\frac{(-1)^{M-k-2}}{2\cos\theta}
\cos (2M-2k-1)\theta
+\frac{1}{2}]  \non \\
&=&
{\cal T}_{\mbox{\scriptsize odd}}^{(k)}.
\eeqa

\section{Matrix element of the second order}
\lab{Section-Matrix-elements-density-operator-second}
In this section, we consider the matrix element
$\bra 0|T_{2}^{(M)}|0\ket$
which contains two $\sigma_{z}$ errors,
defined in Eq.~(\ref{definition-T2M}).
We make good use of the interpretation of creating new modes
discussed in
Section~\ref{Section-physical-interpretation-multi-particle-creation}
for obtaining it.

Let us see the first term of Eq.~(\ref{definition-T2M}).
It suffers two $\sigma_{z}$ errors at the same step as follows,
\beq
\bra 0|(WR_{0})^{2M-k}\sigma_{z}^{(i)}\sigma_{z}^{(j)}
(WR_{0})^{k}W|0\ket
\quad\quad
\mbox{for $1\leq i\leq n$, $1\leq j\leq n$, and $i\neq j$}.
\eeq
Here, we consider the following term,
\beqa
\lefteqn{\bra 0|(WR_{0})^{2(M-k)}\sigma_{z}^{(i)}\sigma_{z}^{(j)}
(WR_{0})^{2k}W|0\ket} \non \\
&=&
(-1)^{M}
[\cos 2(M-k)\theta
\sin (2k+1)\theta \non \\
&&
+\frac{1}{2^{n}-1}
\sin 2(M-k)\theta
\cos (2k+1)\theta
\sum_{x\neq 0}
\sum_{y\neq 0}
\bra x|
\sigma_{z}^{(i)}\sigma_{z}^{(j)}
|y\ket] \non \\
&&\quad\quad
\mbox{for $k=0,1,\cdots,M-1$},
\eeqa
where we use Eqs.~(\ref{Appendix-WR0-2k-W0})
and (\ref{Appendix-R0W-2k-0}) in
Appendixes~\ref{Appendix-subsec-formulas-of-WR0-k-W-0}
and \ref{Appendix-subsec-formulas-of-R0W-k-0}.
To obtain an explicit form of the above,
we have to calculate
$\sum_{x\neq 0}\sum_{y\neq 0}
\bra x|\sigma_{z}^{(i)}\sigma_{z}^{(j)}|y\ket$
for $i\neq j$.

\begin{figure}
\begin{center}
\includegraphics[scale=1.0]{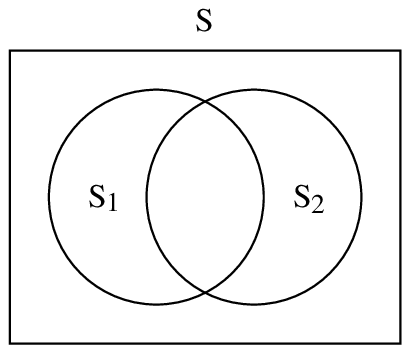}
\end{center}
\caption{Relation between
${\cal S}$, ${\cal S}_{1}$,
and ${\cal S}_{2}$.}
\lab{vector-sets}
\end{figure}

For deriving it,
we define a set
${\cal S}=\{|x\ket: x\in\{0,1\}^{n},x\neq 0\}$,
and its subsets,
${\cal S}_{1}=\{|x\ket: |x\ket\in{\cal S},x_{i}=1\}$
and
${\cal S}_{2}=\{|x\ket: |x\ket\in{\cal S},x_{j}=1\}$,
as shown in Figure~\ref{vector-sets}.
The number of elements of them
and ${\cal S}_{1}\cap{\cal S}_{2}$
are given as
$|{\cal S}|=2^{n}-1$,
$|{\cal S}_{1}|=|{\cal S}_{2}|=2^{n-1}$,
and $|{\cal S}_{1}\cap{\cal S}_{2}|=2^{n-2}$.
A set of basis vectors whose signs are flipped by
$\sigma_{z}^{(i)}\sigma_{z}^{(j)}$ is given by
$({\cal S}_{1}\cup{\cal S}_{2})
\cap
(\overline{{\cal S}_{1}\cap{\cal S}_{2}})$,
and the number of its elements is equal to
$2^{n-1}$.
Hence, the number of elements in ${\cal S}$ whose signs
are not flipped is equal to $(2^{n-1}-1)$.
From these considerations, we obtain
\beq
\sum_{x\neq 0}\sum_{y\neq 0}
\bra x|\sigma_{z}^{(i)}\sigma_{z}^{(j)}|y\ket=-1,
\eeq
and
\beq
\bra 0|(WR_{0})^{2(M-k)}\sigma_{z}^{(i)}\sigma_{z}^{(j)}
(WR_{0})^{2k}W|0\ket
=
{\cal T}_{\mbox{\scriptsize even}}^{(k)}.
\eeq
In a similar way, we can obtain
\beq
\bra 0|(WR_{0})^{2(M-k)-1}\sigma_{z}^{(i)}\sigma_{z}^{(j)}
(WR_{0})^{2k+1}W|0\ket
={\cal T}_{\mbox{\scriptsize odd}}^{(k)}
\quad\quad
\mbox{for $k=0,1,\cdots,M-1$}.
\eeq
Therefore, the matrix element of the first term
in Eq.~(\ref{definition-T2M})
is equal to $[(n-1)/2]\bra 0|T_{1}^{(M)}|0\ket$.

Then, we think about the second term
in Eq.~(\ref{definition-T2M}).
Using Eq.~(\ref{WR0-2l-sigmaZ-WR0-2k-W0}),
we obtain the following term,
\beqa
{\cal T}_{\mbox{\scriptsize even},\mbox{\scriptsize even}}
^{(k,l;\delta_{ij})}
&\equiv&
\bra 0|(WR_{0})^{2(M-k-l)}\sigma_{z}^{(i)}
\biggl[
(WR_{0})^{2l}\sigma_{z}^{(j)}
(WR_{0})^{2k}
W|0\ket
\biggr] \non \\
&=&
\bra 0|(WR_{0})^{2(M-k-l)}\sigma_{z}^{(i)}(WR_{0})^{2(k+l)}W|0\ket \non \\
&&
-\sqrt{2}(-1)^{k}
\frac{\cos(2k+1)\theta}{\cos\theta} \non \\
&&
\times
[-\sqrt{2}\sum_{m=0}^{l-1}\bra 0|(WR_{0})^{2(M-k-l)}
\sigma_{z}^{(i)}(WR_{0})^{2m}W|0\ket
+\bra 0|(WR_{0})^{2(M-k-l)}\sigma_{z}^{(i)}|\eta_{j}\ket].
\lab{definition-T-even-even}
\eeqa
The bracket $[\cdots]$ in the first line
of Eq.~(\ref{definition-T-even-even})
represents that this part is calculated at first.
Here, we use the same technique in
Section~\ref{Section-physical-interpretation-multi-particle-creation}
again.
We expand the matrix element by modes caused by the $\sigma_{z}$ error
and sum up them.
(We eliminate one $\sigma_{z}$ operator
from
${\cal T}_{\mbox{\scriptsize even},\mbox{\scriptsize even}}
^{(k,l;\delta_{ij})}$.)
As a result of Eq.~(\ref{definition-T-even-even}),
we obtain terms which contains only one $\sigma_{z}$ error,
and essentially they have been obtained already in
Sections~\ref{Section-Matrix-elements-density-operator-0th-first}
and \ref{Section-physical-interpretation-multi-particle-creation}.
Here, we introduce a notation of
\beq
{\cal G}^{(1)}(k,l)
=
\bra 0|(WR_{0})^{l}
\sigma_{z}^{(i)}
(WR_{0})^{k}W|0\ket,
\quad\quad
\mbox{for $k=0,1,\cdots$, $l=1,2,\cdots$},
\eeq
and collect its explicit form in
Appendix~\ref{Appendix-subsec-formulas-0-WR0-l-SigmaZ-WR0-k-W0}.
We also collect some formulas of $|\eta_{j}\ket$
and $(|0\ket-|\overline{j}\ket)$
in Appendix~\ref{Appendix-subsec-formulas-0-WR0-k-SigmaZ-eta}.

Using Eqs.~(\ref{Appendix-sum-sin-2l+1-theta}),
(\ref{Appendix-sum-cos-2l+1-theta}),
(\ref{0-WR0-2l-sigmaZ-WR0-2k-W0}),
and (\ref{0-WR0-2k-sigmaZ-eta-j})
in Appendixes~\ref{Appendix-subsec-formulas-for-summation-trigonometric-functions},
\ref{Appendix-subsec-formulas-0-WR0-l-SigmaZ-WR0-k-W0},
and \ref{Appendix-subsec-formulas-0-WR0-k-SigmaZ-eta},
we obtain
\beqa
{\cal T}_{\mbox{\scriptsize even},\mbox{\scriptsize even}}
^{(k,l;\delta_{ij})}
&=&
{\cal G}^{(1)}(2(k+l),2(M-k-l)) \non \\
&&
-\sqrt{2}(-1)^{k}
\frac{\cos(2k+1)\theta}{\cos\theta} \non \\
&&
\times
\{-\sqrt{2}\sum_{m=0}^{l-1}
(-1)^{M-k-l+m}
[\cos 2(M-k-l)\theta \sin(2m+1)\theta \non \\
&&
-\frac{1}{2^{n}-1}\sin 2(M-k-l)\theta\cos(2m+1)\theta] \non \\
&&
-\frac{(-1)^{M-k-l}}{\sqrt{2}}
\frac{\sin 2(M-k-l)\theta}{\cos\theta}\delta_{ij}\} \non \\
&=&
{\cal G}^{(1)}(2(k+l),2(M-k-l)) \non \\
&&
-(-1)^{M}
\frac{\cos(2k+1)\theta}{\cos^{2}\theta} \non \\
&&
\times
\{[\cos 2(M-k-l)\theta \sin 2l\theta
-\frac{1}{2^{n}-1}\sin 2(M-k-l)\theta \cos 2l\theta] \non \\
&&
+(-1)^{-l}(\frac{1}{2^{n}-1}-\delta_{ij})\sin 2(M-k-l)\theta \}.
\lab{Explicit-form-T1-kl-even-even}
\eeqa
From the above, we notice that
${\cal T}_{\mbox{\scriptsize even},\mbox{\scriptsize even}}
^{(k,l;\delta_{ij})}$ depends on
$\delta_{ij}$ (and not on $i$ and $j$).

As results of  similar considerations,
using Eqs.~(\ref{1-phase-error-WR0-2k-W0}),
(\ref{1-phase-error-WR0-(2k+1)-W0}),
(\ref{WR0-2k-eta-i}), (\ref{WR0-2k+1-eta-i}),
and formulas in
Appendixes~\ref{Appendix-subsec-formulas-for-summation-trigonometric-functions},
\ref{Appendix-subsec-formulas-0-WR0-l-SigmaZ-WR0-k-W0},
and \ref{Appendix-subsec-formulas-0-WR0-k-SigmaZ-eta},
we obtain the other matrix elements,
\beqa
{\cal T}_{\mbox{\scriptsize odd},\mbox{\scriptsize even}}
^{(k,l;\delta_{ij})}
&\equiv&
\bra 0|
(WR_{0})^{2(M-k-l)-1}\sigma_{z}^{(i)}
(WR_{0})^{2l}\sigma_{z}^{(j)}
(WR_{0})^{2k+1}
W|0\ket \non \\
&=&
{\cal G}^{(1)}(2(k+l)+1,2(M-k-l)-1) \non \\
&&
+(-1)^{M}
\frac{\sin 2(k+1)\theta}{\cos^{2}\theta} \non \\
&&
\times
\{[\sin (2M-2k-2l-1)\theta \sin 2l\theta
+\frac{1}{2^{n}-1}\cos (2M-2k-2l-1)\theta \cos 2l\theta] \non \\
&&
+(-1)^{l-1}
(\frac{1}{2^{n}-1}-\delta_{ij})\cos (2M-2k-2l-1)\theta \},
\lab{Explicit-form-T1-kl-odd-even} \\
{\cal T}_{\mbox{\scriptsize even},\mbox{\scriptsize odd}}
^{(k,l;\delta_{ij})}
&\equiv&
\bra 0|
(WR_{0})^{2(M-k-l)-1}\sigma_{z}^{(i)}
\biggl[
(WR_{0})^{2l+1}\sigma_{z}^{(j)}
(WR_{0})^{2k}
W|0\ket
\biggr] \non \\
&=&
{\cal G}^{(1)}(2(k+l)+1,2(M-k-l)-1) \non \\
&&
-\sqrt{2}(-1)^{k}
\frac{\cos(2k+1)\theta}{\cos\theta}
[-\sqrt{2}\sum_{m=0}^{l-1}{\cal G}^{(1)}(2m+1,2(M-k-l)-1) \non \\
&&
+\frac{1}{\sqrt{2}}
\bra 0|
(WR_{0})^{2(M-k-l)-1}\sigma_{z}^{(i)}
(|0\ket)-|\overline{j}\ket]\} \non \\
&=&
{\cal G}^{(1)}(2(k+l)+1,2(M-k-l)-1) \non \\
&&
-\sqrt{2}(-1)^{k}
\frac{\cos(2k+1)\theta}{\cos\theta} \non \\
&&
\times
\{-\sqrt{2}\sum_{m=0}^{l-1}
(-1)^{M-k-l-1+m}
[-\sin (2M-2k-2l-1)\theta \cos 2(m+1)\theta \non \\
&&
+\frac{1}{2^{n}-1}\cos (2M-2k-2l-1)\theta \sin 2(m+1)\theta] \non \\
&&
-\frac{(-1)^{M-k-l-1}}{\sqrt{2}}
[\sin (2M-2k-2l-1)\theta
+(-1)^{\delta_{ij}}
\frac{\cos (2M-2k-2l-1)\theta}{\sqrt{2^{n}-1}}]\} \non \\
&=&
{\cal G}^{(1)}(2(k+l)+1,2(M-k-l)-1) \non \\
&&
-(-1)^{M}
\frac{\cos(2k+1)\theta}{\cos^{2}\theta} \non \\
&&
\times
\{[\sin (2M-2k-2l-1)\theta \cos (2l+1)\theta
-\frac{1}{2^{n}-1}\cos (2M-2k-2l-1)\theta \sin (2l+1)\theta] \non \\
&&
+\frac{(-1)^{l}}{\sqrt{2^{n}}}
[\frac{1}{2^{n}-1}+(-1)^{\delta_{ij}}]
\cos (2M-2k-2l-1)\theta \},
\lab{Explicit-form-T1-kl-even-odd} \\
{\cal T}_{\mbox{\scriptsize odd},\mbox{\scriptsize odd}}
^{(k,l;\delta_{ij})}
&\equiv&
\bra 0|
(WR_{0})^{2(M-k-l-1)}\sigma_{z}^{(i)}
(WR_{0})^{2l+1}\sigma_{z}^{(j)}
(WR_{0})^{2k+1}
W|0\ket \non \\
&=&
{\cal G}^{(1)}(2(k+l)+1,2(M-k-l-1)) \non \\
&&
+(-1)^{M}
\frac{\sin 2(k+1)\theta}{\cos^{2}\theta} \non \\
&&
\times
\{-[\cos 2(M-k-l-1)\theta \cos (2l+1)\theta
+\frac{1}{2^{n}-1}\sin 2(M-k-l-1)\theta \sin (2l+1)\theta] \non \\
&&
+\frac{(-1)^{l}}{\sqrt{2^{n}}}
[\frac{1}{2^{n}-1}+(-1)^{\delta_{ij}}]
\sin 2(M-k-l-1)\theta \}.
\lab{Explicit-form-T1-kl-odd-odd}
\eeqa

Finally, we can write $\bra 0|T_{2}^{(M)}|0\ket$ as
\beqa
\bra 0|T_{2}^{(M)}|0\ket
&=&
\frac{n-1}{2}\bra 0|T_{1}^{(M)}|0\ket \non \\
&&
+
\sum_{k=0}^{M-1}\sum_{l=1}^{M-k-1}
[n(n-1)
|{\cal T}_{\mbox{\scriptsize even},\mbox{\scriptsize even}}^{(k,l;0)}|^{2}
+n
|{\cal T}_{\mbox{\scriptsize even},\mbox{\scriptsize even}}^{(k,l;1)}|^{2}] \non \\
&&
+
\sum_{k=0}^{M-1}\sum_{l=1}^{M-k-1}
[n(n-1)
|{\cal T}_{\mbox{\scriptsize odd},\mbox{\scriptsize even}}^{(k,l;0)}|^{2}
+n
|{\cal T}_{\mbox{\scriptsize odd},\mbox{\scriptsize even}}^{(k,l;1)}|^{2}] \non \\
&&
+
\sum_{k=0}^{M-1}\sum_{l=0}^{M-k-1}
[n(n-1)
|{\cal T}_{\mbox{\scriptsize even},\mbox{\scriptsize odd}}^{(k,l;0)}|^{2}
+n
|{\cal T}_{\mbox{\scriptsize even},\mbox{\scriptsize odd}}^{(k,l;1)}|^{2}] \non \\
&&
+
\sum_{k=0}^{M-1}\sum_{l=0}^{M-k-2}
[n(n-1)
|{\cal T}_{\mbox{\scriptsize odd},\mbox{\scriptsize odd}}^{(k,l;0)}|^{2}
+n
|{\cal T}_{\mbox{\scriptsize odd},\mbox{\scriptsize odd}}^{(k,l;1)}|^{2}].
\lab{Explicit-form-matrix-element-T2M}
\eeqa

\section{Large $n$ limit and asymptotic forms of matrix elements}
\lab{Section-large-n-qubits-limit}
The matrix elements,
$\bra 0|T_{1}^{(M)}|0\ket$ and $\bra 0|T_{2}^{(M)}|0\ket$
obtained in
Sections~\ref{Section-Matrix-elements-density-operator-0th-first}
and
\ref{Section-Matrix-elements-density-operator-second},
are too complicated to handle as they are.
In this section, we take the limit of an infinite number of qubits
($n\rightarrow \infty$),
and discuss their asymptotic forms.
We also discuss how to obtain an asymptotic form of
any higher order term under $n\rightarrow \infty$.

We consider the limit of $n\rightarrow \infty$ for $n$-qubit state.
We assume we can take very small $p$,
so that $x=2Mnp$ can be an arbitrary real positive value or $0$.
If $0\leq M<2\pi\sqrt{2^{n}}$,
$M\theta$ converges on a certain value of $\Theta$
($0\leq\Theta<2\pi$) under this limit.
(The definition of $\theta$ is given in Eq.~(\ref{Boyer-parameter-theta}).)
It is reasonable that
we assume $M$ is order of $O(\sqrt{2^{n}})$ or less
and
define
$\Theta\equiv\lim_{n\rightarrow\infty}M\theta$.
Because $\bra 0|T_{h}^{(M)}|0\ket/(Mn)^{h}$
does not depend on $p$,
we can take $n\rightarrow \infty$ for it
naively.
Hence, from Eq.~(\ref{matrix-element-T0}),
we obtain
\beq
\lim_{n\rightarrow \infty}\bra 0|T_{0}^{(M)}|0\ket
=\sin^{2}2\Theta.
\lab{asymptotic-form-T0M}
\eeq

Then, we consider the asymptotic form of $\bra 0|T_{1}^{(M)}|0\ket$
in Eqs.~(\ref{Matrix-element-T1-even}),
(\ref{Matrix-element-T1-odd}),
and (\ref{Explicit-form-matrix-element-T1M}).
To let it converge on finite value,
we divide it by a factor of $Mn$
as shown in Eq.~(\ref{Definitions-p-expanded-density-operators-0123}).
We can obtain
\beq
\lim_{n\rightarrow \infty}
\frac{\bra 0|T_{1}^{(M)}|0\ket}{Mn}
=
\lim_{n\rightarrow \infty}
\frac{1}{M}
\sum_{k=0}^{M-1}
[|\tilde{{\cal T}}_{\mbox{\scriptsize even}}^{(k)}|^{2}
+|\tilde{{\cal T}}_{\mbox{\scriptsize even}}^{(k)}|^{2}],
\lab{asymptotic-form-matrix-element-T1M}
\eeq
where
\beqa
\tilde{{\cal T}}_{\mbox{\scriptsize even}}^{(k)}
&=&
(-1)^{M}\cos 2(M-k)\theta \sin(2k+1)\theta, \non \\
\tilde{{\cal T}}_{\mbox{\scriptsize odd}}^{(k)}
&=&
(-1)^{M}\sin (2M-2k-1)\theta \cos 2(k+1)\theta.
\lab{asymptotic-form-T1M-even-odd}
\eeqa
(We drop the terms with a factor $1/(2^{n}-1)$
in
${\cal T}_{\mbox{\scriptsize even}}^{(k)}$
and
${\cal T}_{\mbox{\scriptsize odd}}^{(k)}$,
and obtain Eq.(\ref{asymptotic-form-T1M-even-odd}).)
We substitute $\Theta=\lim_{n\rightarrow\infty}M\theta$,
$\phi=k\theta$, and
\beq
\lim_{n\rightarrow\infty}
\sum_{k=0}^{M-1}\theta=\int_{0}^{\Theta}d\phi
\eeq
into Eq.~(\ref{asymptotic-form-matrix-element-T1M}),
and obtain
\beqa
\lim_{n\rightarrow \infty}
\frac{\bra 0|T_{1}^{(M)}|0\ket}{Mn}
&=&
\frac{1}{\Theta}
\int_{0}^{\Theta}d\phi
\{[\cos 2(\Theta-\phi)\sin 2\phi]^{2}
+[\sin 2(\Theta-\phi)\cos 2\phi]^{2}\} \non \\
&=&
\frac{1}{2}
-\frac{1}{4}\cos 4\Theta
-\frac{1}{16\Theta}\sin 4\Theta.
\lab{asymptotic-T1M-integral-form}
\eeqa

Next, we consider an asymptotic form of $\bra 0|T_{2}^{(M)}|0\ket$
obtained in
Eqs.~(\ref{Explicit-form-T1-kl-even-even}),
(\ref{Explicit-form-T1-kl-odd-even}),
(\ref{Explicit-form-T1-kl-even-odd}),
(\ref{Explicit-form-T1-kl-odd-odd}),
and (\ref{Explicit-form-matrix-element-T2M}).
Because of convergence, we divide it by $(Mn)^{2}$
as Eq.~(\ref{Definitions-p-expanded-density-operators-0123}).
In the limit of $n\rightarrow\infty$,
we can neglect $\bra 0|T_{1}^{(M)}|0\ket$ and
${\cal T}_{\alpha,\beta}^{(k,l;1)}$ for
$\alpha,\beta\in\{\mbox{even},\mbox{odd}\}$,
and we obtain
\beqa
\lim_{n\rightarrow \infty}
\frac{\bra 0|T_{2}^{(M)}|0\ket}{(Mn)^{2}}
&=&
\lim_{n\rightarrow \infty}
\frac{1}{M^{2}}
[\sum_{k=0}^{M-1}\sum_{l=1}^{M-k-1}
|\tilde{{\cal T}}_{\mbox{\scriptsize even},\mbox{\scriptsize even}}^{(k,l;0)}|^{2}
+
\sum_{k=0}^{M-1}\sum_{l=1}^{M-k-1}
|\tilde{{\cal T}}_{\mbox{\scriptsize odd},\mbox{\scriptsize even}}^{(k,l;0)}|^{2} \non \\
&&
+
\sum_{k=0}^{M-1}\sum_{l=0}^{M-k-1}
|\tilde{{\cal T}}_{\mbox{\scriptsize even},\mbox{\scriptsize odd}}^{(k,l;0)}|^{2}
+
\sum_{k=0}^{M-1}\sum_{l=0}^{M-k-2}
|\tilde{{\cal T}}_{\mbox{\scriptsize odd},\mbox{\scriptsize odd}}^{(k,l;0)}|^{2}],
\lab{Matrix-element-T2M-large-n-limit}
\eeqa
where
\beqa
\tilde{{\cal T}}_{\mbox{\scriptsize even},\mbox{\scriptsize even}}^{(k,l;0)}
&=&
(-1)^{M}[\cos 2(M-k-l)\theta \sin(2k+2l+1)\theta \non \\
&&
-\cos(2k+1)\theta\cos 2(M-k-l)\theta \sin 2l\theta] \non \\
&=&
(-1)^{M}\cos 2(M-k-l)\theta \cos 2l\theta \sin(2k+1)\theta, \non \\
\tilde{{\cal T}}_{\mbox{\scriptsize odd},\mbox{\scriptsize even}}^{(k,l;0)}
&=&
(-1)^{M}\sin (2M-2k-2l-1)\theta \cos 2l\theta \cos 2(k+1)\theta, \non \\
\tilde{{\cal T}}_{\mbox{\scriptsize even},\mbox{\scriptsize odd}}^{(k,l;0)}
&=&
(-1)^{M}[\sin (2M-2k-2l-1)\theta \cos 2(k+l+1)\theta \non \\
&&
-\cos (2k+1)\theta \sin (2M-2k-2l-1)\theta \cos (2l+1)\theta] \non \\
&=&
(-1)^{M+1}\sin (2M-2k-2l-1)\theta \sin (2l+1)\theta \sin (2k+1)\theta, \non \\
\tilde{{\cal T}}_{\mbox{\scriptsize odd},\mbox{\scriptsize odd}}^{(k,l;0)}
&=&
(-1)^{M}\cos 2(M-k-l-1)\theta \sin (2l+1)\theta \cos 2(k+1)\theta.
\lab{asymptotic-form-T2M-even-odd}
\eeqa
(We use $\lim_{n\rightarrow\infty}\cos\theta=1$
because of Eq.~(\ref{Boyer-parameter-theta}).)
This asymptotic form contains only the terms
where $\sigma_{z}$ errors occur at different steps
and at different qubits
from each other.
Hence, defining $\varphi=l\theta$ and
$\lim_{n\rightarrow\infty}\sum_{l=0}^{M-k}\theta=\int_{0}^{\Theta-\phi}d\varphi$,
we obtain
\beqa
\lim_{n\rightarrow \infty}
\frac{\bra 0|T_{2}^{(M)}|0\ket}{(Mn)^{2}}
&=&
\frac{1}{\Theta^{2}}
\int_{0}^{\Theta}d\phi
\int_{0}^{\Theta-\phi}d\varphi \non \\
&&
\{[\cos 2(\Theta-\phi-\varphi)\cos 2\varphi \sin 2\phi]^{2} \non \\
&&
+[\sin 2(\Theta-\phi-\varphi)\cos 2\varphi \cos 2\phi]^{2} \non \\
&&
+[\sin 2(\Theta-\phi-\varphi)\sin 2\varphi \sin 2\phi]^{2} \non \\
&&
+[\cos 2(\Theta-\phi-\varphi)\sin 2\varphi \cos 2\phi]^{2}\} \non \\
&=&
\frac{1}{4}
-\frac{1}{16}\cos 4\Theta
-\frac{3}{64\Theta}\sin 4\Theta.
\lab{asymptotic-T2M-integral-form}
\eeqa

Seeing Eqs.~(\ref{asymptotic-form-T1M-even-odd}),
(\ref{asymptotic-T1M-integral-form}),
(\ref{asymptotic-form-T2M-even-odd}),
(\ref{asymptotic-T2M-integral-form}),
and formulas of
Appendix~\ref{Appendix-subsec-formulas-0-WR0-l-SigmaZ-WR0-k-W0},
we find how to obtain the asymptotic form of
$h$-th density operator
($h=1,2,\cdots$)
under $n\rightarrow \infty$.
We derive it in
Appendixes~\ref{Appendix-derivation-asymptotic-forms-matrix-elements},
\ref{Appendix-formulas-of-0-WR0-lk-WR0-l1-sigmaZ-eta-i},
and \ref{Appendix-formulas-of-0-WR0-sigmaZ-sigmaZ-(0-i0)}.
Here, we use only its result.
Preparing an $h$-digit binary string
$\alpha=(\alpha_{1},\cdots,\alpha_{h})\in\{0,1\}^{h}$,
we define the following $2^{h}$ terms,
\beqa
\lefteqn{|\tilde{{\cal T}}_{\alpha_{1},\cdots,\alpha_{h}}
(\phi_{1},\cdots,\phi_{h})|^{2}} \non \\
&=&
[
{\sin\brace{\cos}}_{\alpha_{1}}(2\phi_{1})
{\cos\brace{\sin}}_{\alpha_{2}}(2\phi_{2})
\cdots
{\cos\brace{\sin}}_{\alpha_{h}}(2\phi_{h})
{\cos\brace{\sin}}_{\oplus_{s=1}^{h}\alpha_{s}}
(2(\Theta-\sum_{s=1}^{h}\phi_{s}))
]^{2} \non \\
&&
\quad\quad
\mbox{for $h=1,2,\cdots$},
\lab{Diagrammatic-rule}
\eeqa
where
\beq
{f\brace{g}}_{\alpha}(x)
\equiv
\left\{
\begin{array}{lll}
f(x) & \mbox{for $\alpha=0$} \\
g(x) & \mbox{for $\alpha=1$}
\end{array}
\right..
\eeq
We notice that
the function of $\phi_{1}$ and the other functions of
$\phi_{2}$, $\cdots$, $\phi_{h}$,
$\Theta-\sum_{s=1}^{h}\phi_{s}$
are different
(sine and cosine functions are put in reverse).
These terms are integrated as
\beqa
\lefteqn{\lim_{n\rightarrow \infty}
\frac{\bra 0|T_{h}^{(M)}|0\ket}{(Mn)^{h}}} \non \\
&=&
\frac{1}{\Theta^{h}}
\int_{0}^{\Theta}d\phi_{1}
\int_{0}^{\Theta-\phi_{1}}d\phi_{2}
\cdots
\int_{0}^{\Theta-\phi_{1}-\cdots-\phi_{h-1}}d\phi_{h}
\sum_{(\alpha_{1},\cdots,\alpha_{h})\in\{0,1\}^{h}}
|\tilde{{\cal T}}_{\alpha_{1},\cdots,\alpha_{h}}
(\phi_{1},\cdots,\phi_{h})|^{2}.
\lab{Diagrammatic-rule-integral}
\eeqa
We can find that the expression of
Eqs.~(\ref{Diagrammatic-rule}) and (\ref{Diagrammatic-rule-integral})
is coincide with
Eqs.~(\ref{asymptotic-T1M-integral-form})
and
(\ref{asymptotic-T2M-integral-form}).

Here, let us calculate the asymptotic form of $\bra 0|T_{3}^{(M)}|0\ket$.
From the above rules,
we obtain
\beqa
\lim_{n\rightarrow \infty}
\frac{\bra 0|T_{3}^{(M)}|0\ket}{(Mn)^{3}}
&=&
\frac{1}{\Theta^{3}}
\int_{0}^{\Theta}d\phi_{1}
\int_{0}^{\Theta-\phi_{1}}d\phi_{2}
\int_{0}^{\Theta-\phi_{1}-\phi_{2}}d\phi_{3} \non \\
&&
\{
[\sin 2\phi_{1}\cos 2\phi_{2}\cos 2\phi_{3}
\cos 2(\Theta-\phi_{1}-\phi_{2}-\phi_{3})]^{2} \non \\
&&
+[\cos 2\phi_{1}\cos 2\phi_{2}\cos 2\phi_{3}
\sin 2(\Theta-\phi_{1}-\phi_{2}-\phi_{3})]^{2} \non \\
&&
+[\sin 2\phi_{1}\sin 2\phi_{2}\cos 2\phi_{3}
\sin 2(\Theta-\phi_{1}-\phi_{2}-\phi_{3})]^{2} \non \\
&&
+[\cos 2\phi_{1}\sin 2\phi_{2}\cos 2\phi_{3}
\cos 2(\Theta-\phi_{1}-\phi_{2}-\phi_{3})]^{2} \non \\
&&
+[\sin 2\phi_{1}\cos 2\phi_{2}\sin 2\phi_{3}
\sin 2(\Theta-\phi_{1}-\phi_{2}-\phi_{3})]^{2} \non \\
&&
+[\cos 2\phi_{1}\cos 2\phi_{2}\sin 2\phi_{3
}\cos 2(\Theta-\phi_{1}-\phi_{2}-\phi_{3})]^{2} \non \\
&&
+[\sin 2\phi_{1}\sin 2\phi_{2}\sin 2\phi_{3}
\cos 2(\Theta-\phi_{1}-\phi_{2}-\phi_{3})]^{2} \non \\
&&
+[\cos 2\phi_{1}\sin 2\phi_{2}\sin 2\phi_{3}
\sin 2(\Theta-\phi_{1}-\phi_{2}-\phi_{3})]^{2}\} \non \\
&=&
\frac{1}{12}
+\frac{3-16\Theta^{2}}{1536\Theta^{2}}\cos 4\Theta
-\frac{1+32\Theta^{2}}{2048\Theta^{3}}\sin 4\Theta.
\lab{asymptotic-T3M-integral-form}
\eeqa

We pay attention to the following facts.
If we expand the asymptotic forms of
Eqs.~(\ref{asymptotic-form-T0M}),
(\ref{asymptotic-T1M-integral-form}),
(\ref{asymptotic-T2M-integral-form}),
and (\ref{asymptotic-T3M-integral-form})
in powers of $\Theta$,
we obtain
\beq
\lim_{n\rightarrow \infty}
\frac{\bra 0|T_{h}^{(M)}|0\ket}{(Mn)^{h}}
=
\left\{
\begin{array}{ll}
4\Theta^{2}+O(\Theta^{3})      & \mbox{for $h=0$} \\
(8/3)\Theta^{2}+O(\Theta^{3})  & \mbox{for $h=1$} \\
\Theta^{2}+O(\Theta^{3})       & \mbox{for $h=2$} \\
(4/15)\Theta^{2}+O(\Theta^{3}) & \mbox{for $h=3$}
\end{array}
\right..
\eeq
Hence, they converge to $0$ under the limit of
$\Theta\rightarrow 0$ (or $M\rightarrow 0$).
This means that the probability of observing $|0\ket$
for the uniform superposition is almost $0$,
and it is reasonable.

\section{Numerical calculations of physical quantities}
\lab{Section-numerical-calculations}
In this section, we carry out numerical calculation of
$\bra 0|\rho^{(M)}|0\ket$
for the asymptotic form
under the $n\rightarrow \infty$ limit,
and investigate physical quantities explained in
Section~\ref{section-perturbative-method}.
Especially, we discuss the critical point
$x_{\mbox{\scriptsize c}}$,
over which the quantum algorithm comes not to be available
for the threshold probability $P_{\mbox{\scriptsize th}}$.

In the perturbation theory,
we can rewrite Eq.~(\ref{density-operator-expansion-p-powers})
under the limit of $n\rightarrow\infty$ as
\beqa
P_{\mbox{\scriptsize rob}}(\Theta,x)
&\equiv &
\lim_{n\rightarrow\infty}
\bra 0|\rho^{(M)}|0\ket \non \\
&=&
C_{0}(\Theta)+C_{1}(\Theta)x+\frac{1}{2}C_{2}(\Theta)x^{2}+\cdots \non \\
&=&
\sum_{h=0}^{\infty}C_{h}(\Theta)\frac{1}{h!}x^{h},
\lab{matrix-element-asymptotic-expansion}
\eeqa
where
$\Theta=\lim_{n\rightarrow\infty}M\theta$,
$x=2Mnp$,
\beqa
C_{0}(\Theta)&=&F_{0}(\Theta), \non \\
C_{1}(\Theta)&=&-F_{0}(\Theta)+\frac{1}{2}F_{1}(\Theta), \non \\
C_{h}(\Theta)
&=&
(-1)^{h}\sum_{j=0}^{h}
(-\frac{1}{2})^{j}
\frac{h!}{(h-j)!}
F_{j}(\Theta)
\quad\quad
\mbox{for $h=0,1,\cdots$},
\lab{coefficients-matrix-element-asymptotic-expansion}
\eeqa
and
\beq
F_{h}(\Theta)
=\lim_{n\rightarrow \infty}
\frac{\bra 0|T_{h}^{(M)}|0\ket}{(Mn)^{h}}
\quad\quad
\mbox{for $h=0,1,\cdots$}.
\lab{Definition-F-h-Theta}
\eeq
$F_{h}(\Theta)$ for $h=0,1,2,3$ are obtained in
Eqs.~(\ref{asymptotic-form-T0M}),
(\ref{asymptotic-T1M-integral-form}),
(\ref{asymptotic-T2M-integral-form}),
and (\ref{asymptotic-T3M-integral-form}).
Using the rules of
Eqs.~(\ref{Diagrammatic-rule})
and (\ref{Diagrammatic-rule-integral}),
we obtain higher order terms in
Appendix~\ref{Appendix-notes-for-numerical-calculations}.

The original model that we define in Section~\ref{Section-model-decoherence}
has two independent parameters,
$M$ and $p$,
and $n$ takes a fixed finite value.
On the other hand,
the representation of Eq.~(\ref{matrix-element-asymptotic-expansion})
has $\Theta$ and $x$ as independent parameters,
and $n$ gets infinity,
so that it does not have a certain fixed value for $n$.
(It is clear that $\Theta$ and $x$ are independent of each other
from their definitions.)
These difference reflects physical quantities of
$M_{\mbox{\scriptsize th}}(p,P_{\mbox{\scriptsize th}})$,
$p_{\mbox{\scriptsize c}}$,
$\tilde{M}_{\mbox{\scriptsize th}}(x,P_{\mbox{\scriptsize th}})$,
and $x_{\mbox{\scriptsize c}}$,
which are introduced in Section~\ref{section-perturbative-method}.
When we estimate these quantities numerically,
we examine their meanings and differences.

In this section,
we make numerical calculations up to the fifth order correction.
To investigate the range of $x$ where our perturbative
approach is valid,
we need to estimate the sixth order term of
Eq.~(\ref{matrix-element-asymptotic-expansion}).
From the discussion
in Appendix~\ref{Appendix-notes-for-numerical-calculations},
we can consider it is reliable around $0\leq x\leq 1.35$,
so that the sixth order correction of $P_{\mbox{\scriptsize rob}}(\Theta, x)$
is bounded to $10^{-3}$.
We compare the perturbation theory with results of
Monte Carlo simulations of our model,
and confirm its reliability in Figures~\ref{numcal-probx} and \ref{numcal-ProbTh}.
In these simulations,
setting $n=8$ ($8$ qubits),
we fix $p$ and cause $\sigma_{z}$ errors at random
on each trial.
We take an average of
$\bra 0| \rho^{(M)}|0\ket_{p}$,
the probability of observing $|0\ket$ at the $M$-th step
($M=0,1,\cdots,M_{\mbox{\scriptsize max}}(=12)$),
with $20000$ trials for each certain value of $p$.

In Figures~\ref{numcal-Mth050p} and \ref{numcal-Mth050x},
we consider perturbations up to an odd order
(the first, third, and fifth).
If we sum up even number of correction terms,
$\bra 0| \rho^{(M)}|0\ket$ with fixed $\Theta$ (or $M$)
does not decrease monotonously against
$x=2Mnp$ (or $p$).
It turns for increasing from some value of $x$ (or $p$),
so that sometimes we cannot find $p_{\mbox{\scriptsize c}}$
(or $x_{\mbox{\scriptsize c}}$)
by numerical calculation.
Hence, we always consider corrections up to an odd order.

\begin{figure}
\begin{center}
\includegraphics[scale=1.0]{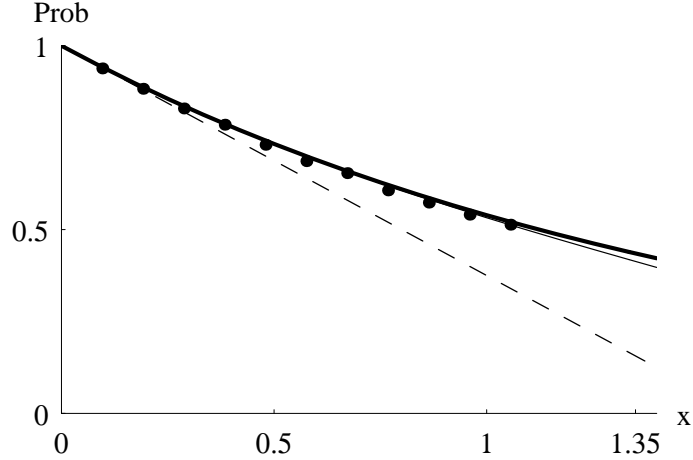}
\end{center}
\caption{Variation of
$P_{\mbox{\scriptsize rob}}(\Theta, x)$
against $x$ with fixed $\Theta=\pi/4$.
(It means $M$ is fixed to $M_{\mbox{\scriptsize max}}=(\pi/4)\sqrt{2^{n}}$.)
A thin dashed curve, a thin solid curve, and a thick solid curve show
perturbations up to the first, third, and fifth
order each.
Black circles represent results obtained by Monte Carlo simulations
of $n=8$ case ($8$ qubits) with $M_{\mbox{\scriptsize max}}=12$.
Each circle is obtained for
$x=2M_{\mbox{\scriptsize max}}np=192 p$,
where $p$ is varied from
$5\times 10^{-4}$ to $5.5\times 10^{-3}$
at interval of $5\times 10^{-4}$.
In these simulations,
we make $20000$ trials for taking an average.}
\lab{numcal-probx}
\end{figure}

Figure~\ref{numcal-probx} shows a variation of
$P_{\mbox{\scriptsize rob}}(\Theta, x)$
against $x$ with fixed $\Theta=\pi/4$,
namely
$M$ is fixed to $M_{\mbox{\scriptsize max}}=(\pi/4)\sqrt{2^{n}}$.
(Hence, the independent parameter is only $p$ actually,
but $n$ is infinite.)
We draw the curves of Eq.~(\ref{matrix-element-asymptotic-expansion})
up to the first, third, and fifth order corrections,
and plot simulation results.
At $x=0$, there is no error and
$P_{\mbox{\scriptsize rob}}$ is equal to unity.
As the error rate $x$ gets larger,
$P_{\mbox{\scriptsize rob}}$ decreases monotonously.

\begin{figure}
\begin{center}
\includegraphics[scale=1.0]{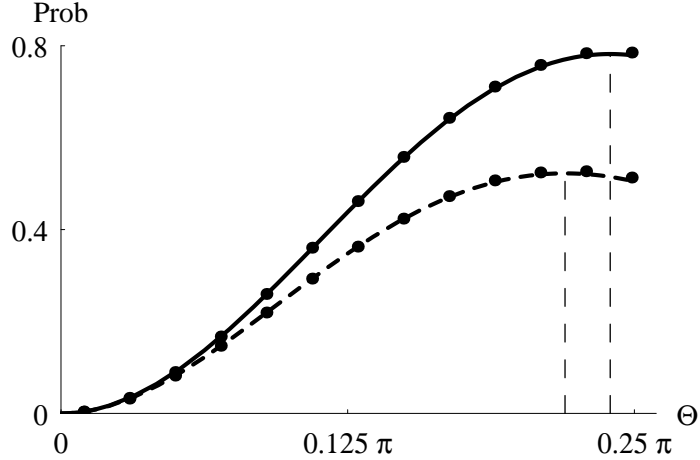}
\end{center}
\caption{Variation of
$P_{\mbox{\scriptsize rob}}(\Theta, x)$
(perturbation up to the fifth correction)
against $\Theta$ with fixed $p$.
The thick solid curve and the thick dashed curve represent
$p=2.0\times 10^{-3}$ and
$p=5.5\times 10^{-3}$ each, for $n=8$.
Black circles are results of Monte Carlo simulations.
These two curves show
$\Theta_{\mbox{\scriptsize th}}(p_{\mbox{\scriptsize c}})$
is lower than $\pi/4$.
It is confirmed by the simulation results.}
\lab{numcal-ProbTh}
\end{figure}

Figure~\ref{numcal-ProbTh} shows a variations of
$P_{\mbox{\scriptsize rob}}(\Theta, x)$
against $\Theta$ with fixed $p$.
Because we use the variable $x=2Mnp$ instead of $p$
in the perturbation theory,
we have to rewrite
\beq
x=2Mnp=2(n\sqrt{2^{n}})\Theta p,
\eeq
and give some finite $n$.
In Figure~\ref{numcal-ProbTh},
we set $n=8$ and draw curves of perturbation theory
up to the fifth order
against $\Theta$ with fixed $p$
($p=2.0\times 10^{-3}$ for the thick solid curve,
and $p=5.5\times 10^{-3}$ for the thick dashed curve).
We also plot results of the simulations.

From Figure~\ref{numcal-ProbTh},
we notice that the maximum value of $P_{\mbox{\scriptsize rob}}$
is taken at $\Theta<\pi/4$ for each $p$
(we show these points with vertical thin dashed lines),
and the shift gets larger as $p$ increases.
It means
$\Theta_{\mbox{\scriptsize th}}
(p_{\mbox{\scriptsize c}};
P_{\mbox{\scriptsize th}})$
gets smaller than $\pi/4$,
as $P_{\mbox{\scriptsize th}}$ decreases.
(We write
$\Theta_{\mbox{\scriptsize th}}(p;
P_{\mbox{\scriptsize th}})
\equiv\lim_{n\rightarrow\infty}
M_{\mbox{\scriptsize th}}(p;
P_{\mbox{\scriptsize th}})\theta$,
and
$M_{\mbox{\scriptsize th}}(p;
P_{\mbox{\scriptsize th}})$ means the least number of the operations
iterated for amplifying the probability
of $|0\ket$ to
$P_{\mbox{\scriptsize th}}$
under the error rate $p$.)

\begin{figure}
\begin{center}
\includegraphics[scale=1.0]{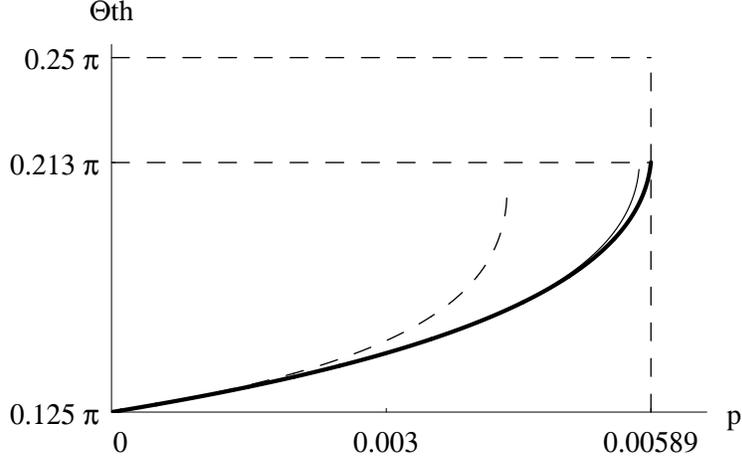}
\end{center}
\caption{Variation of
$\Theta_{\mbox{\scriptsize th}}(p)$
against $p$ with fixed $n(=8)$ for $P_{\mbox{\scriptsize th}}=1/2$.
A thin dashed curve, a thin solid curve, and a thick solid curve show
perturbations up to the first, third, and fifth
order each.
The algorithm cannot observe $|0\ket$ with the probability of
$1/2$ or more for $p > 0.00589$.
Hence,
$p_{\mbox{\scriptsize c}}\approx 0.00589$
is the critical point of $P_{\mbox{\scriptsize th}}=1/2$.
We obtain 
$\Theta_{\mbox{\scriptsize th}}(p_{\mbox{\scriptsize c}})
\approx 0.213 \pi<\pi/4$.}
\lab{numcal-Mth050p}
\end{figure}

Then, let us see the behaviour of the algorithm with fixing the threshold
of the probability on $P_{\mbox{\scriptsize th}}$.
Figure~\ref{numcal-Mth050p} represents
a variation of $\Theta_{\mbox{\scriptsize th}}(p)$ against $p$
with $n=8$ ($8$ qubits)
for $P_{\mbox{\scriptsize th}}=1/2$.
Seeing it,
we can confirm that $\Theta_{\mbox{\scriptsize th}}(p)$
cannot reach to $\pi/4$,
even if $p=p_{\mbox{\scriptsize c}}$.
It is consistent with results of Figure~\ref{numcal-ProbTh}.

\begin{figure}
\begin{center}
\includegraphics[scale=1.0]{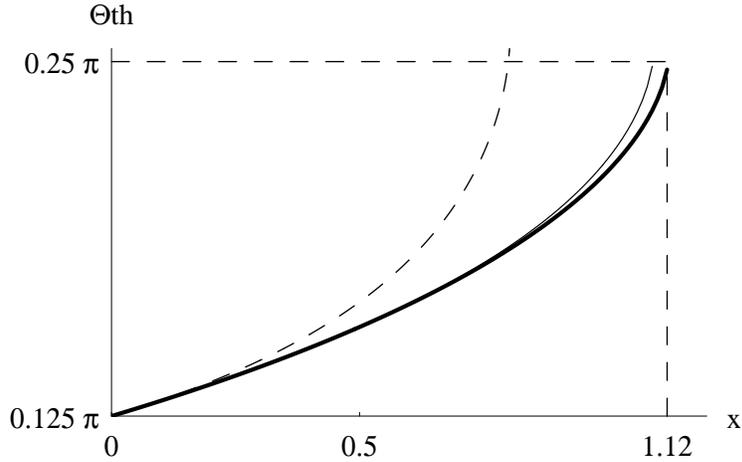}
\end{center}
\caption{Variation of
$\tilde{\Theta}_{\mbox{\scriptsize th}}(x)$
against $x$ with $P_{\mbox{\scriptsize th}}=1/2$.
A thin dashed line, a thin solid line, and a thick solid line show
perturbations up to the first, third, and fifth
order each.
The algorithm cannot observe $|0\ket$ with the probability of
$1/2$ or more for $x> 1.12$.
Hence,
$x_{\mbox{\scriptsize c}}\approx 1.12$
is the critical point of $P_{\mbox{\scriptsize th}}=1/2$.
We obtain
$\tilde{\Theta}_{\mbox{\scriptsize th}}(x_{\mbox{\scriptsize c}})
\approx 0.247\pi$.}
\lab{numcal-Mth050x}
\end{figure}

When we draw curves of Figures~\ref{numcal-ProbTh} and \ref{numcal-Mth050p},
we have to put finite positive $n$.
(We set $n$ on $8$.)
This treatment cannot be fully justified,
because Eq.~(\ref{matrix-element-asymptotic-expansion})
is obtained with the $n\rightarrow \infty$
limit.
Next, we compute physical quantities with taking independent parameters $\Theta$ and $x$.
(We need not give finite $n$.)
Figure~\ref{numcal-Mth050x} shows a variation of $\tilde{\Theta}_{\mbox{\scriptsize th}}(x)$
against $x$ with $P_{\mbox{\scriptsize th}}=1/2$,
where
$\tilde{\Theta}_{\mbox{\scriptsize th}}(x;P_{\mbox{\scriptsize th}})
\equiv\lim_{n\rightarrow\infty}
\tilde{M}_{\mbox{\scriptsize th}}(x;P_{\mbox{\scriptsize th}})\theta$.
($\tilde{M}_{\mbox{\scriptsize th}}(x;P_{\mbox{\scriptsize th}})$
represents the least number of the operations
to amplify the probability of $|0\ket$ to $P_{\mbox{\scriptsize th}}$
under given $x$.)
Seeing Figure~\ref{numcal-Mth050x},
we find 
$\tilde{\Theta}_{\mbox{\scriptsize th}}(x)$
increases as $x$ gets larger from $x=0$,
and it reaches to the maximum value at $x=x_{\mbox{\scriptsize c}}$.
Comparing Figures~\ref{numcal-Mth050p} and \ref{numcal-Mth050x},
we notice
$\tilde{\Theta}_{\mbox{\scriptsize th}}(x_{\mbox{\scriptsize c}})
>\Theta_{\mbox{\scriptsize th}}(p_{\mbox{\scriptsize c}})$
for $P_{\mbox{\scriptsize th}}=1/2$.
(We can actually confirm
$\tilde{\Theta}_{\mbox{\scriptsize th}}(x_{\mbox{\scriptsize c}})
\geq
\Theta_{\mbox{\scriptsize th}}(p_{\mbox{\scriptsize c}})$
for $0< \forall P_{\mbox{\scriptsize th}}\leq 1$
by numerical calculations.)

\begin{figure}
\begin{center}
\includegraphics[scale=1.0]{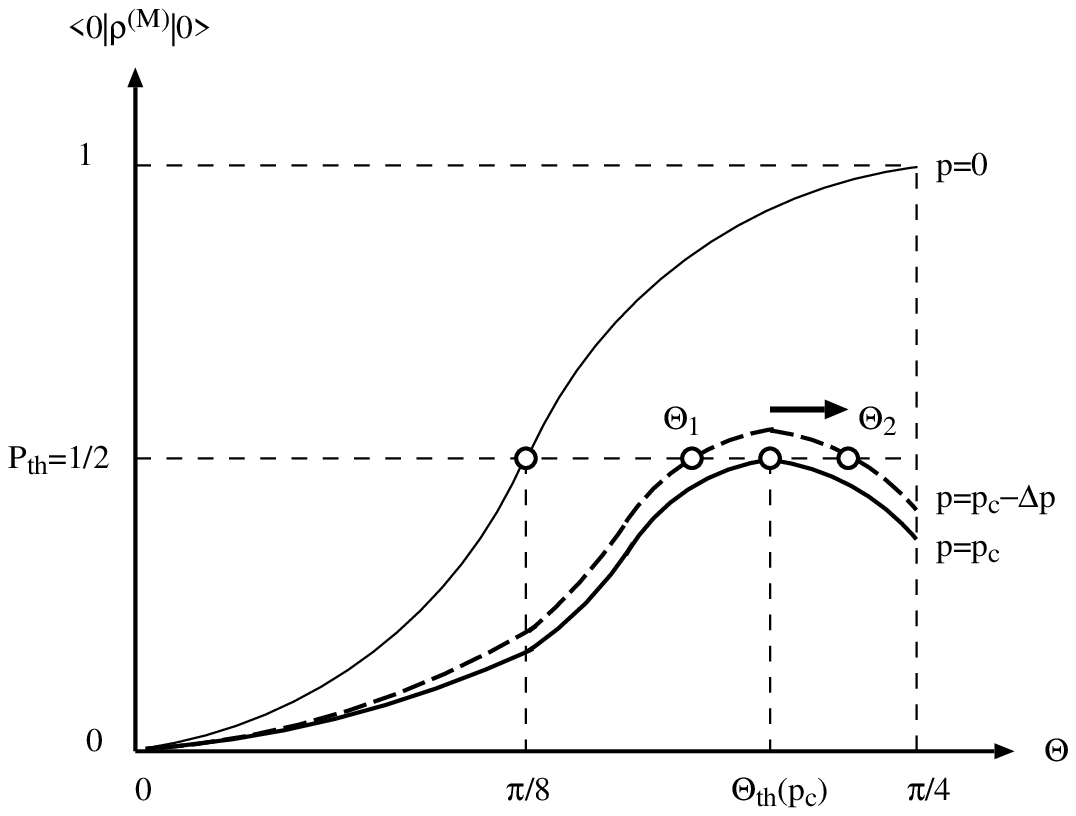}
\end{center}
\caption{Variation of
$\bra 0|\rho^{(M)}|0\ket$
against $\Theta$ with $0\leq p\leq p_{\mbox{\scriptsize c}}$.
The threshold of probability is set to $P_{\mbox{\scriptsize th}}=1/2$.
White circles represent
$\Theta_{\mbox{\scriptsize th}}(p)$
and 
$\tilde{\Theta}_{\mbox{\scriptsize th}}(x)$.}
\lab{probth-px}
\end{figure}

It can be explained as  follows.
Figure~\ref{probth-px} illustrates a variation of
$\bra 0|\rho^{(M)}|0\ket$
against $\Theta$ for $0\leq p\leq p_{\mbox{\scriptsize c}}$.
In the case of $p=0$ (no decoherence),
we can obtain
$\tilde{\Theta}_{\mbox{\scriptsize th}}(x=0)
=\Theta_{\mbox{\scriptsize th}}(p=0)
=\pi/8$
for $P_{\mbox{\scriptsize th}}=1/2$.
(If $p=0$, we obtain $x=2Mnp=0$.)
If we let $p$ get larger until $p=p_{\mbox{\scriptsize c}}$,
$\Theta_{\mbox{\scriptsize th}}(p)$ increases gradually.
Through this process,
both $x$ and $\tilde{\Theta}_{\mbox{\scriptsize th}}(x)$
increase.
When we reach at $p=p_{\mbox{\scriptsize c}}$,
we obtain
$\tilde{\Theta}_{\mbox{\scriptsize th}}(x)
=\Theta_{\mbox{\scriptsize th}}(p_{\mbox{\scriptsize c}})
<\pi/4$
for $x=2M_{\mbox{\scriptsize th}}(p)np_{\mbox{\scriptsize c}}
<x_{\mbox{\scriptsize c}}$.

Although $p$ gets the allowed maximum value of $p_{\mbox{\scriptsize c}}$,
we want to increase both $x$ and $\tilde{\Theta}_{\mbox{\scriptsize th}}(x)$
still more.
To increase $x$,
we take the following trick.
We decrease $p$ by infinitesimal $\Delta p$,
as shown in Figure~\ref{probth-px}.
Then, $\bra 0|\rho^{(M)}|0\ket$
takes $P_{\mbox{\scriptsize th}}=1/2$
at two points of $\Theta$, and we write them as
$\Theta_{1}$ and $\Theta_{2}$
($\Theta_{1}<\Theta_{2}$).
At this time,
we take the large one of them as $\tilde{\Theta}_{\mbox{\scriptsize th}}(x)$,
so that $\tilde{\Theta}_{\mbox{\scriptsize th}}(x)=\Theta_{2}$.

Because $\bra 0|\rho^{(M)}|0\ket$
takes the local minimum value at
$\Theta_{\mbox{\scriptsize th}}(p_{\mbox{\scriptsize c}})$
for $p=p_{\mbox{\scriptsize c}}$,
we can obtain
\beq
\left.
\frac{\partial \bra 0|\rho^{(M)}|0\ket}
{\partial \Theta}
\right|_{
{
\Theta=\Theta_{\mbox{\tiny th}}(p_{\mbox{\tiny c}})
}
\atop
{
p=p_{\mbox{\tiny c}}
}
}
=
\left.
\frac{\partial \bra 0|\rho^{(M)}|0\ket}
{\partial \Theta_{\mbox{\scriptsize th}}(p)}
\right|_{p=p_{\mbox{\tiny c}}}
=
\left.
\frac{\partial \bra 0|\rho^{(M)}|0\ket}
{\partial p}
\frac{\partial p}
{\partial \Theta_{\mbox{\scriptsize th}}(p)}
\right|_{p=p_{\mbox{\tiny c}}}
=0,
\eeq
\beq
\left.
\frac{\partial \bra 0|\rho^{(M)}|0\ket}
{\partial p}
\right|_{p=p_{\mbox{\tiny c}}}
\neq 0,
\eeq
and
\beq
\left.
\frac{\partial p}
{\partial \Theta_{\mbox{\scriptsize th}}(p)}
\right|_{p=p_{\mbox{\tiny c}}}
=0
\quad
\mbox{or}
\quad
\left.
\frac{\partial \Theta_{\mbox{\scriptsize th}}(p)}
{\partial p}
\right|_{p=p_{\mbox{\tiny c}}}
=-\infty.
\eeq
Hence, the difference of
$[\Theta_{2}-\Theta_{\mbox{\scriptsize th}}(p_{\mbox{\scriptsize c}})]$
can be quite large,
and
$\Theta_{2}(p_{\mbox{\scriptsize c}}-\Delta p)
\geq
\Theta_{\mbox{\scriptsize th}}(p_{\mbox{\scriptsize c}})
p_{\mbox{\scriptsize c}}$ is possible.
Therefore,
we can make
$\tilde{\Theta}_{\mbox{\scriptsize th}}(x)
\geq
\Theta_{\mbox{\scriptsize th}}(p_{\mbox{\scriptsize c}})$
and
$x
\geq
2M_{\mbox{\scriptsize th}}(p_{\mbox{\scriptsize c}})np_{\mbox{\scriptsize c}}$.
(These considerations can be applied to
$0< \forall
P_{\mbox{\scriptsize th}}\leq 1$ as well.)

\begin{figure}
\begin{center}
\includegraphics[scale=1.0]{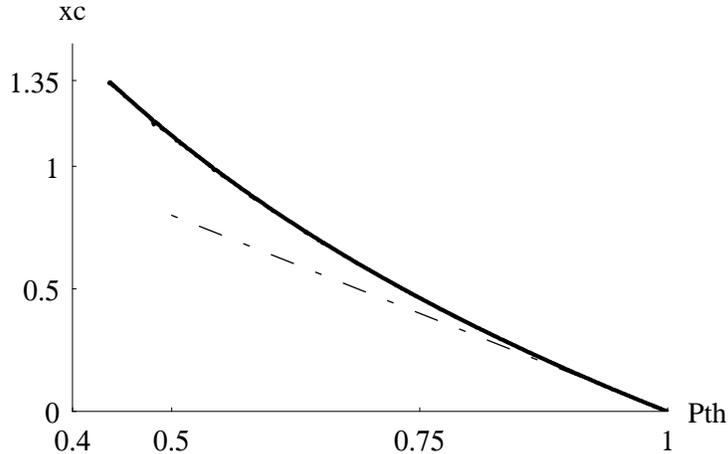}
\end{center}
\caption{Variation of
$x_{\mbox{\scriptsize c}}$
against $P_{\mbox{\scriptsize th}}$.
A thick solid curve represents perturbation up to the fifth order.
A thin dashed line shows its tangent at $P_{\mbox{\scriptsize th}}=1$
given by Eq.~(\ref{tangent-line}).}
\lab{numcal-xcPth}
\end{figure}

Then, we move on to the variation of
$x_{\mbox{\scriptsize c}}$ against $P_{\mbox{\scriptsize th}}$,
which is shown in Figure~\ref{numcal-xcPth}.
We obtain it as follows.
We calculate $\tilde{\Theta}_{\mbox{\scriptsize th}}(x)$
for given $P_{\mbox{\scriptsize th}}$
as varying $x$ from $0$.
(We use the Newton's method for obtaining a root of $\Theta$
for the equation of
$P_{\mbox{\scriptsize rob}}(\Theta,x)
=P_{\mbox{\scriptsize th}}$
with given $x$.)
When $x$ gets a certain value,
we cannot find a root for $\tilde{\Theta}_{\mbox{\scriptsize th}}(x)$,
and we regard it as $x_{\mbox{\scriptsize c}}$.
By repeating this calculation,
we obtain the curve of Figure~\ref{numcal-xcPth}.

In these calculations,
we notice
$\tilde{\Theta}_{\mbox{\scriptsize th}}
(x_{\mbox{\scriptsize c}};P_{\mbox{\scriptsize th}})
\simeq\pi/4$
for the range of $x$ and $P_{\mbox{\scriptsize th}}$
where the perturbation theory is reliable
($0\leq x\leq 1.35$).
It is caused by the approximately symmetric property of
$F_{h}(\Theta)$
obtained in Section~\ref{Section-large-n-qubits-limit}
and Appendix~\ref{Appendix-notes-for-numerical-calculations}
as
\beq
F_{h}((\pi/4)+\Delta)
\simeq
F_{h}((\pi/4)-\Delta)
\quad\quad
\mbox{for $0<\Delta\ll (\pi/4)$ and $h=0,1,\cdots$}.
\eeq
However, strictly speaking,
$\tilde{\Theta}_{\mbox{\scriptsize th}}
(x_{\mbox{\scriptsize c}};P_{\mbox{\scriptsize th}})$
cannot be a constant for $\forall x_{\mbox{\scriptsize c}}$
and $\forall P_{\mbox{\scriptsize th}}$.

Using Eq.~(\ref{matrix-element-asymptotic-expansion}),
a tangent at $P_{\mbox{\scriptsize th}}=1$
is given by
\beq
x_{\mbox{\scriptsize c}}
=
c(1-P_{\mbox{\scriptsize th}}),
\quad\quad
c=-\frac{1}{C_{1}(\pi/4)}=\frac{8}{5},
\lab{tangent-line}
\eeq
because
$\tilde{\Theta}_{\mbox{\scriptsize th}}(x_{\mbox{\scriptsize c}})=\pi/4$
and
$x_{\mbox{\scriptsize c}}=0$
for $P_{\mbox{\scriptsize th}}=1$.
It means that
the algorithm is available
for
$2Mnp<(8/5)(1-P_{\mbox{\scriptsize th}})$
around $P_{\mbox{\scriptsize th}}\simeq 1$,
and this relation approximately holds for a wide range of
$P_{\mbox{\scriptsize th}}$.
This result is similar to a work
obtained by E.~Bernstein and U.~Vazirani \cite{Bernstein}.
We mention it in Section~\ref{subsection-Accuracy-of-quantum-gates}.

Figure~\ref{numcal-xcPth} shows a transition
about whether quantum computing
is available or not
for threshold probability
$P_{\mbox{\scriptsize th}}$.
Here, let us consider where is a classical searching
on the phase diagram of Figure~\ref{numcal-xcPth}.
We assume that we are looking for one item among unsorted $2^{n}$ items.
If we examine $M$ items from them
in a classical manner,
we can find it with probability
$P=M/2^{n}$.
Now, let us regard $M$ as the number of the quantum operations iterated
and
$P$ as the threshold
$P_{\mbox{\scriptsize th}}$
for the algorithm in classical regime.
If we give $P_{\mbox{\scriptsize th}}>0$
(and it is not infinitesimal),
the classical searching takes $M\simeq O(2^{n})$
and $x\gg x_{\mbox{\scriptsize c}}$.
Hence, it is located far away upward in the non-available region of
the quantum algorithm in Figure~\ref{numcal-xcPth}.

On the other hand,
if we consider the neighbourhood of
$P_{\mbox{\scriptsize th}}=0$,
it becomes subtle.
The classical searching can take small $M$,
and it can approach to the available region of
the quantum algorithm.
Furthermore,
in the limit of $P_{\mbox{\scriptsize th}}\rightarrow 0$,
we can expect
$M_{\mbox{\scriptsize th}}(p_{\mbox{\scriptsize c}})
\rightarrow 0$
(but $n\rightarrow\infty$)
for the quantum algorithm,
so that
behaviour of $x_{\mbox{\scriptsize c}}$
in the neighbourhood of $P_{\mbox{\scriptsize th}}=0$
might be singular.

From these discussions,
we consider that
a quantum to classical phase transition
of the algorithm is described around $P_{\mbox{\scriptsize th}}\simeq 0$
in Figure~\ref{numcal-xcPth}.
We cannot say anything
about it by our approach,
because $x_{\mbox{\scriptsize c}}$
for $P_{\mbox{\scriptsize th}}\simeq 0$
is outside the domain where the perturbation theory is reliable.

\section{Discussions}
\lab{Section-discussions}
In this section,
we think about related work obtained by
E.~Bernstein and U.~Vazirani,
and how the phase error is caused.
Then, we give other discussions about our results.

\subsection{Accuracy of quantum gates}
\lab{subsection-Accuracy-of-quantum-gates}
E.~Bernstein and U.~Vazirani consider accuracy of quantum gates
for quantum computation \cite{Bernstein}\cite{Preskill}.
Let us think about a quantum computer which is designed to apply
$T$ unitary transformations,
$U_{1},\cdots,U_{T}$,
in succession ($T$ steps) to the initial state $|\phi_{0}\ket$, as follows,
\beq
|\phi_{0}\ket
\stackrel{U_{1}}{\longrightarrow}
|\phi_{1}\ket
\stackrel{U_{2}}{\longrightarrow}
\cdots
\stackrel{U_{T}}{\longrightarrow}
|\phi_{T}\ket,
\eeq
so that $|\phi_{t}\ket=U_{t}|\phi_{t-1}\ket$
and
$\bra \phi_{t}|\phi_{t}\ket
=\bra \phi_{0}|\phi_{0}\ket
=1$
for $t=1,\cdots,T$.
On the other hand, we assume that it actually applies
$\tilde{U}_{t}$
which is slightly different from $U_{t}$
to the state because of incomplete accuracy,
\beq
|\phi_{0}\ket
\stackrel{\tilde{U}_{1}}{\longrightarrow}
|\tilde{\phi}_{1}\ket
\stackrel{\tilde{U}_{2}}{\longrightarrow}
\cdots
\stackrel{\tilde{U}_{T}}{\longrightarrow}
|\tilde{\phi}_{T}\ket,
\lab{unitary-error-model}
\eeq
so that
$|\tilde{\phi}_{t}\ket=\tilde{U}_{t}|\tilde{\phi}_{t-1}\ket$,
$|\tilde{\phi}_{0}\ket=|\phi_{0}\ket$,
and
$\bra \tilde{\phi}_{t}|\tilde{\phi}_{t}\ket=1$
for $t=1,\cdots,T$.
(We are considering errors of unitary transformations,
and it does not cause dissipation to the quantum computer.)

Defining unnormalised states
\beq
|E_{t}\ket=(\tilde{U}_{t}-U_{t})|\phi_{t-1}\ket,
\lab{difference-from-unitary-error}
\eeq
we obtain
\beqa
|\tilde{\phi}_{1}\ket
&=&
|\phi_{1}\ket+|E_{1}\ket, \non \\
|\tilde{\phi}_{2}\ket
&=&
|\phi_{2}\ket+|E_{2}\ket+\tilde{U}_{2}|E_{1}\ket, \non \\
\vdots && \non \\
|\tilde{\phi}_{T}\ket
&=&
|\phi_{T}\ket+|E_{T}\ket+\tilde{U}_{T}|E_{T-1}\ket
\cdots
+
\tilde{U}_{T}\tilde{U}_{T-1}\cdots\tilde{U}_{2}|E_{1}\ket,
\eeqa
and
\beq
\bra\phi_{T}|\tilde{\phi}_{T}\ket
\geq
1-\sum_{t=1}^{T}\sqrt{\bra E_{t}|E_{t}\ket}.
\eeq

Here, we assume that the error of each step is bounded as
\beq
\sqrt{\bra E_{t}|E_{t}\ket}\leq\epsilon
\quad\quad
\mbox{for $t=1,\cdots,T$}.
\eeq
We obtain
\beq
|\bra\phi_{T}|\tilde{\phi}_{T}\ket|^{2}
\geq
(1-T\epsilon)^{2}
=
1-2T\epsilon+O(\epsilon^{2}).
\lab{probability-unitary-error-model}
\eeq
Hence, if the error of the unitary transformation at each step
is bounded to $\epsilon$,
the probability of detecting $|\phi_{T}\ket$
that we want as the final state
is at least $(1-2T\epsilon)$.

On the other hand,
from Eq.~(\ref{difference-from-unitary-error}),
we obtain
\beq
\bra \phi_{t}|\tilde{U}_{T}|\phi_{t-1}\ket
=
1+\bra \phi_{t}|E_{t}\ket
\geq
1-\epsilon,
\eeq
and we can rewrite it with density operator as
\beq
\mbox{tr}(\rho\rho')\geq 1-2\epsilon+O(\epsilon^{2}),
\eeq
where $\rho=|\phi_{t}\ket\bra\phi_{t}|$
and $\rho'=\tilde{U}_{T}|\phi_{t}\ket\bra\phi_{t}|\tilde{U}_{T}^{\dagger}$.
In this paper,
we consider the decoherence defined in
Eq.~(\ref{one-qubit-phase-error}).
Although it is different from the error of unitary transformations
in Eq.~(\ref{unitary-error-model}),
we can obtain
\beq
\mbox{tr}(\rho\rho')\geq 1-p,
\eeq
and regard it as inaccuracy of operation for
$\epsilon=p/2$ at each step.

If we require $|\bra\phi_{T}|\tilde{\phi}_{T}\ket|^{2}> P_{\mbox{\scriptsize th}}$
for a threshold of the probability
that the quantum computer gives a correct answer,
we can obtain
\beq
1-P_{\mbox{\scriptsize th}}
> 1-|\bra\phi_{T}|\tilde{\phi}_{T}\ket|^{2}
\simeq 2T\epsilon,
\eeq
as the first order estimation.
Substituting $\epsilon=p/2$, and $T=2Mn$
which is the number of quantum gates during the whole process
(the number of decoherences caused)
into the above,
we can obtain $2Mnp< 1-P_{\mbox{\scriptsize th}}$.
This is similar to the result
obtained in Section~\ref{Section-numerical-calculations},
except for a factor.

\subsection{How the phase error occurs}
\lab{Subsection-how-the-phase-error-occurs}
We give a mechanism which  causes the phase error of
Eq.~(\ref{one-qubit-phase-error})
for an instance.
We can think this error to be quite possible for proposed implementations
of quantum computation \cite{Cirac-Gershenfeld}.
Let us consider two spin-$1/2$ systems described as $3$-component
normalised vectors of
$\mbox{\boldmath $\sigma$}^{A}$ (qubit) and
$\mbox{\boldmath $\sigma$}^{E}$ (environment),
whose interaction is given by their inner product of
$\kappa \mbox{\boldmath $\sigma$}^{A}\cdot \mbox{\boldmath $\sigma$}^{E}$.

If there is weak external magnetic field
along $z$-direction
$\mbox{\boldmath $B$}=(0,0,B_{z})$,
both of them align themselves with
$z$-direction,
so that
$\mbox{\boldmath $\sigma$}^{A}=(0,0,\pm 1)$
and
$\mbox{\boldmath $\sigma$}^{E}=(0,0,\pm 1)$.
Hence, we obtain an effective Hamiltonian of
$\Delta H\simeq\kappa \sigma_{z}^{A}\sigma_{z}^{E}$,
and a time-evolution operator
\beqa
U_{\Delta}^{AE}
&=&
\exp (-i\frac{t}{\hbar}\Delta H) \non \\
&=&
\bordermatrix{
       & \bra 00|      & \bra 01|    & \bra 10|    & \bra 11|     \cr
|00\ket & e^{-i\theta} & 0           & 0           & 0            \cr
|01\ket & 0            & e^{i\theta} & 0           & 0            \cr
|10\ket & 0            & 0           & e^{i\theta} & 0            \cr
|11\ket & 0            & 0           & 0           & e^{-i\theta} \cr
},
\eeqa
on the logical basis
$|ij\ket=|i^{A}\ket|i^{E}\ket$ for $i,j\in\{0,1\}$,
where $\theta=(\kappa/\hbar)t$.

We assume that the initial state of the systems $A$ and $E$ is given as
$|\varphi^{A}\ket|+^{E}\ket$,
where $|\varphi^{A}\ket$ is an arbitrary state of $A$
and
$|+^{E}\ket=(1/\sqrt{2})(|0^{E}\ket+i|1^{E}\ket)$.
It evolves as follows \cite{Schumacher},
\beqa
\rho^{A}=|\varphi^{A}\ket\bra\varphi^{A}|
\rightarrow\rho'^{A}
&=&
\mbox{tr}_{E}
[U_{\Delta}^{AE}
(|\varphi^{A}\ket\bra\varphi^{A}|
\otimes
|+^{E}\ket\bra +^{E}|)
U_{\Delta}^{AE}{}^{\dagger}] \non \\
&=&
\sum_{\mu\in\{+,-\}}M_{\mu}^{A}\rho^{A}M_{\mu}^{A}{}^{\dagger},
\eeqa
where
\beq
M_{\mu}^{A}=\bra\mu^{E}|U_{\Delta}^{AE}|+^{E}\ket,
\eeq
and
\beq
|\pm^{E}\ket=\frac{1}{\sqrt{2}}(|0^{E}\ket\pm i|1^{E}\ket).
\eeq

Here, we introduce a $2\times 2$ unitary transformation,
\beqa
V=V^{\dagger}
=\frac{1}{\sqrt{2}}
\bordermatrix{
       & \bra 0| & \bra 1| \cr
|0\ket & 1       & 1       \cr
|1\ket & i       & -i      \cr
},
\quad\quad
VV^{\dagger}=\mbox{\boldmath $I$},
\eeqa
and we obtain
\beq
|+^{E}\ket=V^{E}|0^{E}\ket,
\quad\quad
|-^{E}\ket=V^{E}|1^{E}\ket.
\eeq
We can rewrite $M_{\mu}$ as
\beqa
M_{+}^{A}&=&\bra 0^{E}|V^{E}{}^{\dagger}U_{\Delta}^{AE}V^{E}|0^{E}\ket, \non \\
M_{-}^{A}&=&\bra 1^{E}|V^{E}{}^{\dagger}U_{\Delta}^{AE}V^{E}|0^{E}\ket.
\lab{matrix-M-Kraus-representation}
\eeqa
The matrix of Eq.~(\ref{matrix-M-Kraus-representation})
is given by
\beqa
V^{E}{}^{\dagger}U_{\Delta}^{AE}V^{E}
&=&
\left[
\renewcommand{\arraystretch}{1.2}
\begin{array}{c|c}
V^{\dagger} & 0           \\
\hline
0           & V^{\dagger}
\end{array}
\right]
\left[
\renewcommand{\arraystretch}{1.2}
\begin{array}{c}
\smash{\lower2.0ex\hbox{$U_{\Delta}^{AE}$}} \\
\\
\end{array}
\right]
\left[
\renewcommand{\arraystretch}{1.2}
\begin{array}{c|c}
V & 0 \\
\hline
0 & V
\end{array}
\right] \non \\
&=&
\left[
\renewcommand{\arraystretch}{1.2}
\begin{array}{c|c}
R^{\dagger} & 0 \\
\hline
0           & R
\end{array}
\right],
\eeqa
on the logical basis, where
\beq
R
=
\left(
\begin{array}{cc}
\cos\theta  & i\sin\theta \\
i\sin\theta & \cos\theta
\end{array}
\right).
\eeq

Hence, we obtain
\beq
M_{+}=\cos\theta\mbox{\boldmath $I$}^{A}
\quad\quad
M_{-}=-i\sin\theta\sigma_{z}^{A}.
\eeq
Therefore,
the time evolution of the system $A$ is described as
\beq
\rho^{A}
\rightarrow
\rho'^{A}=\cos^{2}\theta\rho^{A}+\sin^{2}\theta\sigma_{z}\rho^{A}\sigma_{z},
\eeq
and it is equivalent to Eq.~(\ref{one-qubit-phase-error}).

If we assume that each qubit of the quantum computer
interacts with an external spin-$1/2$ particle
under the weak magnetic field every time interval,
our model can give a reasonable description of its decoherence.

\subsection{Other discussions}
\lab{Subsection-other-discussions}
From Figure~\ref{numcal-xcPth},
we find
$x_{\mbox{\scriptsize c}}=
2\tilde{M}_{\mbox{\scriptsize th}}(x_{\mbox{\scriptsize c}})np\simeq O(1)$
for suitable threshold probability
($1/2\leq P_{\mbox{\scriptsize th}}\leq 1$, for example).
It means that if the error ratio $p$ is smaller
than an inverse of the number of quantum gates
$(2Mn)^{-1}$,
the algorithm is reliable.
If this observation holds good for other quantum algorithms,
it can serve a strong foundation to realize quantum computation.
We cannot investigate a quantum to classical phase transition of
the algorithm,
because it is outside the reliable domain of our perturbation theory.
For studying it precisely,
we may need to construct an exact solvable model of a quantum system with
decoherence.

\bigskip
\noindent
{\bf \large Acknowledgements}
\smallskip

We thank D.~K.~L.~Oi and A.~T.~Costa, Jr. for helpful comments
about Section~\ref{Section-numerical-calculations}.
We also thank A.~K.~Ekert for encouragements.

\appendix
\section{Formulas for deriving matrix elements of density operators}
\lab{Appendix-formulas-for-calculating-matrix-elements}
In this section, we collect some formulas that are used for
deriving the matrix element
of the density operator $T_{h}^{(M)}$.

\subsection{Formulas of $(WR_{0})^{k}W|0\ket$}
\lab{Appendix-subsec-formulas-of-WR0-k-W-0}
We derive an explicit form of $(WR_{0})^{k}W|0\ket$,
where $|0\ket$ is an $n$-qubit ($n\geq 2$) initial state of
$|0\ket\otimes \cdots \otimes|0\ket$.
Let us think an $n$-qubit state of
\beq
|\Psi\ket
=
a_{0}|0\ket
+
a_{1}
\sum_{{x\neq 0} \atop {x\in\{0,1\}^{n}}}|x\ket.
\eeq
Using
\beq
W|x\ket
=
\frac{1}{\sqrt{2^{n}}}\sum_{y\in\{0,1\}^{n}}
(-1)^{x\cdot y}|y\ket
\quad\quad
\mbox{for $\forall x\in\{0,1\}^{n}$},
\eeq
where
$x\cdot y$ represents an inner product of
$n$-digit binary strings of $x,y\in\{0,1\}^{n}$,
that is
$x\cdot y=\sum_{i=1}^{n}x_{i}y_{i}$,
we can derive
\beqa
WR_{0}|\Psi\ket
&=&
W(-a_{0}|0\ket+a_{1}\sum_{x\neq 0}|x\ket) \non \\
&=&
\frac{1}{\sqrt{2^{n}}}
[-a_{0}\sum_{y}|y\ket
+a_{1}\sum_{x\neq 0}\sum_{y}(-1)^{x\cdot y}|y\ket] \non \\
&=&
\frac{1}{\sqrt{2^{n}}}
[(-a_{0}+(2^{n}-1)a_{1})|0\ket
-(a_{0}+a_{1})\sum_{y\neq 0}|y\ket].
\lab{Appendix-WR0Psi}
\eeqa

Then we introduce a parameter $\theta$
to simplify notations of states \cite{Boyer-Brassard},
\beq
\sin\theta=\frac{1}{\sqrt{2^{n}}},
\quad\quad
\cos\theta=\sqrt{\frac{2^{n}-1}{2^{n}}}.
\lab{Appendix-parameter-theta}
\eeq
Using $\theta$, we obtain the following trigonometric formulas,
\beqa
\frac{1}{\sqrt{2^{n}}}
[\mp\sin\varphi+(2^{n}-1)\frac{\cos\varphi}{\sqrt{2^{n}-1}}]
&=&
\mp\sin\theta\sin\varphi+\cos\theta\cos\varphi \non \\
&=&
\cos(\varphi\pm\theta),
\lab{Appendix-trigonometric1} \\
\frac{1}{\sqrt{2^{n}}}
[\mp\cos\varphi-(2^{n}-1)\frac{\sin\varphi}{\sqrt{2^{n}-1}}]
&=&
-\sin(\varphi\pm\theta),
\lab{Appendix-trigonometric2} \\
\frac{1}{\sqrt{2^{n}}}
(\pm\sin\varphi+\frac{\cos\varphi}{\sqrt{2^{n}-1}})
&=&
\frac{1}{\sqrt{2^{n}-1}}
(\pm\cos\theta\sin\varphi+\sin\theta\cos\varphi) \non \\
&=&
\pm\frac{\sin(\varphi\pm\theta)}{\sqrt{2^{n}-1}},
\lab{Appendix-trigonometric3} \\
\frac{1}{\sqrt{2^{n}}}
(\pm\cos\varphi-\frac{\sin\varphi}{\sqrt{2^{n}-1}})
&=&
\pm\frac{\cos(\varphi\pm\theta)}{\sqrt{2^{n}-1}}.
\lab{Appendix-trigonometric4}
\eeqa

From these relations,
we can obtain the following results,
\beqa
W|0\ket
&=&
\frac{1}{\sqrt{2^{n}}}
\sum_{x}|x\ket
=
\sin\theta|0\ket
+
\frac{\cos\theta}{\sqrt{2^{n}-1}}
\sum_{x\neq 0}|x\ket, \non \\
(WR_{0})W|0\ket
&=&
\cos 2\theta|0\ket
-
\frac{\sin 2\theta}{\sqrt{2^{n}-1}}
\sum_{x\neq 0}|x\ket.
\eeqa
In general, we obtain
\beqa
(WR_{0})^{2k}W|0\ket
&=&
(-1)^{k}
[\sin(2k+1)\theta|0\ket
+
\frac{\cos(2k+1)\theta}{\sqrt{2^{n}-1}}
\sum_{x\neq 0}|x\ket]
\quad\quad
\mbox{for $k=0,1,\cdots$},
\lab{Appendix-WR0-2k-W0} \\
(WR_{0})^{2k+1}W|0\ket
&=&
(-1)^{k}
[\cos 2(k+1)\theta|0\ket
-
\frac{\sin 2(k+1)\theta}{\sqrt{2^{n}-1}}
\sum_{x\neq 0}|x\ket]
\quad\quad
\mbox{for $k=0,1,\cdots$}.
\lab{Appendix-WR0-(2k+1)-W0}
\eeqa

\subsection{Formulas of $(R_{0}W)^{k}|0\ket$}
\lab{Appendix-subsec-formulas-of-R0W-k-0}
From formulas obtained in
Appendix~\ref{Appendix-subsec-formulas-of-WR0-k-W-0}
we obtain
\beqa
(R_{0}W)^{2k}|0\ket
&=&
W(WR_{0})^{2k}W|0\ket \non \\
&=&
\frac{(-1)^{k}}{\sqrt{2^{n}}}
[\sin(2k+1)\theta \sum_{x}|x\ket
+
\frac{\cos(2k+1)\theta}{\sqrt{2^{n}-1}}
(2^{n}|0\ket-\sum_{x}|x\ket)] \non \\
&=&
\frac{(-1)^{k}}{\sqrt{2^{n}}}
\{[\sin(2k+1)\theta
+(2^{n}-1)\frac{\cos(2k+1)\theta}{\sqrt{2^{n}-1}}]|0\ket \non \\
&&\quad\quad
+[\sin(2k+1)\theta -\frac{\cos(2k+1)\theta}{\sqrt{2^{n}-1}}]
\sum_{x\neq 0}|x\ket\} \non \\
&=&
(-1)^{k}
[\cos 2k\theta |0\ket
+
\frac{\sin 2k\theta}{\sqrt{2^{n}-1}}\sum_{x\neq 0}|x\ket]
\quad\quad
\mbox{for $k=0,1,\cdots$},
\lab{Appendix-R0W-2k-0} \\
(R_{0}W)^{2k+1}|0\ket
&=&
(-1)^{k}
[-\sin (2k+1)\theta |0\ket
+
\frac{\cos (2k+1)\theta}{\sqrt{2^{n}-1}}\sum_{x\neq 0}|x\ket]
\quad\quad
\mbox{for $k=0,1,\cdots$}.
\lab{Appendix-R0W-(2k+1)-0}
\eeqa

\subsection{Formulas for summation of trigonometric functions}
\lab{Appendix-subsec-formulas-for-summation-trigonometric-functions}
When we calculate the matrix element of $\bra 0|T_{h}^{(M)}|0\ket$,
we often have to sum up trigonometric functions.
In this paper, we use the following four formulas,
which can be proved by the inductive method \cite{Mangulis},
\beqa
\sum_{l=0}^{n}(-1)^{l}\sin(2l+1)\theta
&=&
\frac{(-1)^{n}}{2}
\frac{\sin 2(n+1)\theta}{\cos\theta}
\quad\quad
\mbox{for $n=0,1,\cdots$},
\lab{Appendix-sum-sin-2l+1-theta} \\
\sum_{l=0}^{n}(-1)^{l}\sin 2(l+1)\theta
&=&
\frac{\sin\theta}{2\cos\theta}
+\frac{(-1)^{n}}{2\cos\theta}
\sin (2n+3)\theta
\quad\quad
\mbox{for $n=0,1,\cdots$},
\lab{Appendix-sum-sin-2(l+1)-theta} \\
\sum_{l=0}^{n}(-1)^{l}\cos(2l+1)\theta
&=&
\frac{1+(-1)^{n}\cos 2(n+1)\theta}{2\cos\theta}
\quad\quad
\mbox{for $n=0,1,\cdots$},
\lab{Appendix-sum-cos-2l+1-theta} \\
\sum_{l=0}^{n}(-1)^{l}\cos 2(l+1)\theta
&=&
\frac{1}{2}
+\frac{(-1)^{n}}{2\cos\theta}
\cos(2n+3)\theta
\quad\quad
\mbox{for $n=0,1,\cdots$}.
\lab{Appendix-sum-cos-2(l+1)-theta}
\eeqa

\subsection{Formulas of
${\cal G}^{(1)}(k,l)=
\bra 0|(WR_{0})^{l}\sigma_{z}^{(i)}(WR_{0})^{k}W|0\ket$}
\lab{Appendix-subsec-formulas-0-WR0-l-SigmaZ-WR0-k-W0}
From Eqs.~(\ref{Appendix-WR0-2k-W0}), (\ref{Appendix-WR0-(2k+1)-W0}),
(\ref{Appendix-R0W-2k-0}), and (\ref{Appendix-R0W-(2k+1)-0}),
we can derive the following formulas,
\beqa
{\cal G}^{(1)}(2k,2l)
&=&
\bra 0|(WR_{0})^{2l}\sigma_{z}^{(i)}(WR_{0})^{2k}W|0\ket \non \\
&=&
(-1)^{l+k}
[\cos 2l\theta \bra 0|
+
\frac{\sin 2l\theta}{\sqrt{2^{n}-1}}\sum_{x\neq 0}\bra x|] \non \\
&&
\times
\sigma_{z}^{(i)}
[\sin(2k+1)\theta|0\ket
+
\frac{\cos(2k+1)\theta}{\sqrt{2^{n}-1}}
\sum_{y\neq 0}|y\ket] \non \\
&=&
(-1)^{l+k}
[\cos 2l\theta \sin(2k+1)\theta
-\frac{1}{2^{n}-1}\sin 2l\theta\cos(2k+1)\theta]\non \\
&&\quad\quad
\mbox{for $k=0,1,\cdots$, $l=1,2,\cdots$},
\lab{0-WR0-2l-sigmaZ-WR0-2k-W0} \\
{\cal G}^{(1)}(2k,2l+1)
&=&
\bra 0|(WR_{0})^{2l+1}\sigma_{z}^{(i)}(WR_{0})^{2k}W|0\ket \non \\
&=&
(-1)^{l+k}
[-\sin (2l+1)\theta \sin(2k+1)\theta
-\frac{1}{2^{n}-1}\cos (2l+1)\theta\cos(2k+1)\theta] \non \\
&&\quad\quad
\mbox{for $k=0,1,\cdots$, $l=0,1,\cdots$},
\lab{0-WR0-(2l+1)-sigmaZ-WR0-2k-W0} \\
{\cal G}^{(1)}(2k+1,2l)
&=&
\bra 0|(WR_{0})^{2l}\sigma_{z}^{(i)}(WR_{0})^{2k+1}W|0\ket \non \\
&=&
(-1)^{l+k}
[\cos 2l\theta \cos 2(k+1)\theta
+\frac{1}{2^{n}-1}\sin 2l\theta\sin 2(k+1)\theta] \non \\
&&\quad\quad
\mbox{for $k=0,1,\cdots$, $l=1,2,\cdots$},
\lab{0-WR0-2l-sigmaZ-WR0-(2k+1)-W0} \\
{\cal G}^{(1)}(2k+1,2l+1)
&=&
\bra 0|(WR_{0})^{2l+1}\sigma_{z}^{(i)}(WR_{0})^{2k+1}W|0\ket \non \\
&=&
(-1)^{l+k}
[-\sin (2l+1)\theta \cos 2(k+1)\theta
+\frac{1}{2^{n}-1}\cos (2l+1)\theta\sin 2(k+1)\theta] \non \\
&&\quad\quad
\mbox{for $k=0,1,\cdots$, $l=0,1,\cdots$},
\lab{0-WR0-(2l+1)-sigmaZ-WR0-(2k+1)-W0}
\eeqa

\subsection{Formulas of $\bra 0|(WR_{0})^{k}\sigma_{z}^{(i)}|\eta_{j}\ket$
and
$(1/\sqrt{2})\bra 0|(WR_{0})^{k}\sigma_{z}^{(i)}(|0\ket-|\overline{j}\ket)$}
\lab{Appendix-subsec-formulas-0-WR0-k-SigmaZ-eta}
From Eqs.~(\ref{definition-eta-i}), (\ref{definition-overline-i})
in Section~\ref{Section-physical-interpretation-multi-particle-creation},
and Eqs.~(\ref{Appendix-R0W-2k-0}), (\ref{Appendix-R0W-(2k+1)-0})
in Appendix~\ref{Appendix-subsec-formulas-of-R0W-k-0},
we can derive the following formulas,
\beqa
\bra 0|(WR_{0})^{2k}\sigma_{z}^{(i)}|\eta_{j}\ket
&=&
(-1)^{k}
[\cos 2k\theta \bra 0|
+
\frac{\sin 2k\theta}{\sqrt{2^{n}-1}}\sum_{x\neq 0}\bra x|]
\sigma_{z}^{(i)}
\frac{1}{\sqrt{2^{n-1}}}
\sum_{y_{j}=1}|y\ket \non \\
&=&
(-1)^{k}
\frac{\sin 2k\theta}{\sqrt{2^{n}-1}\sqrt{2^{n-1}}}
\delta_{ij}(-1)2^{n-1} \non \\
&=&
-(-1)^{k}
\sqrt{\frac{2^{n-1}}{2^{n}-1}}\sin 2k\theta
\delta_{ij} \non \\
&=&
-\frac{(-1)^{k}}{\sqrt{2}}
\frac{\sin 2k\theta}{\cos\theta}
\delta_{ij},
\lab{0-WR0-2k-sigmaZ-eta-j} \\
\bra 0|(WR_{0})^{2k+1}\sigma_{z}^{(i)}|\eta_{j}\ket
&=&
-\frac{(-1)^{k}}{\sqrt{2}}
\frac{\cos (2k+1)\theta}{\cos\theta}
\delta_{ij},
\lab{0-WR0-(2k+1)-sigmaZ-eta-j} \\
\frac{1}{\sqrt{2}}
\bra 0|(WR_{0})^{2k}\sigma_{z}^{(i)}(|0\ket-|\overline{j}\ket)
&=&
\frac{1}{\sqrt{2}}
(-1)^{k}
[\cos 2k\theta \bra 0|
+
\frac{\sin 2k\theta}{\sqrt{2^{n}-1}}\sum_{x\neq 0}\bra x|]
\sigma_{z}^{(i)}
(|0\ket-|\overline{j}\ket) \non \\
&=&
\frac{(-1)^{k}}{\sqrt{2}}
[\cos 2k\theta
-(-1)^{\delta_{ij}}
\frac{\sin 2k\theta}{\sqrt{2^{n}-1}}],
\lab{0-WR0-2k-sigmaZ-0-j} \\
\frac{1}{\sqrt{2}}
\bra 0|(WR_{0})^{2k+1}\sigma_{z}^{(i)}(|0\ket-|\overline{j}\ket)
&=&
-\frac{(-1)^{k}}{\sqrt{2}}
[\sin (2k+1)\theta
+(-1)^{\delta_{ij}}
\frac{\cos (2k+1)\theta}{\sqrt{2^{n}-1}}].
\lab{0-WR0-(2k+1)-sigmaZ-0-j}
\eeqa

\subsection{Derivation of the asymptotic forms of matrix elements}
\lab{Appendix-derivation-asymptotic-forms-matrix-elements}
In this section, we derive the asymptotic forms of matrix elements,
Eqs.~(\ref{Diagrammatic-rule})
and (\ref{Diagrammatic-rule-integral}) introduced
in Section~\ref{Section-large-n-qubits-limit}.

Comparing Eqs.~(\ref{Explicit-form-matrix-element-T2M})
and (\ref{Matrix-element-T2M-large-n-limit}),
we notice that contributions of $\bra 0|T_{h}^{(M)}|0\ket/(Mn)^{h}$
under $n\rightarrow\infty$
comes from only terms in which $\sigma_{z}$ errors
occur at different steps and different qubits.
The reason is as follows.
The term of
\beq
|\bra 0|[\mbox{product of $(WR_{0})$s and $\sigma_{z}$s}]|0\ket|^{2}
\eeq
never exceeds unity,
and the number of terms where $\sigma_{z}$ errors occur
at the same step or the same qubit in $\bra 0|T_{h}^{(M)}|0\ket$
is at most $O((2M)^{h}n^{h-1})$.
Because we divide $\bra 0|T_{h}^{(M)}|0\ket$ by $(Mn)^{h}$,
they are eliminated under $n\rightarrow\infty$,
as far as $h$ is finite.

\begin{figure}
\begin{center}
\includegraphics[scale=1.0]{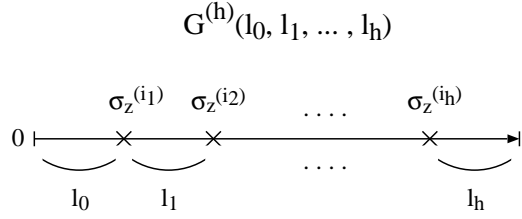}
\end{center}
\caption{Diagram of $h$ errors,
${\cal G}^{(h)}(l_{0},l_{1},\cdots,l_{h})$.}
\lab{error-diagram}
\end{figure}

An asymptotic form of the matrix element with $h$ errors
at different steps and qubits,
as shown in Figure~\ref{error-diagram},
are given by
\beqa
\lefteqn{
\lim_{n\rightarrow\infty}
{\cal G}^{(h)}(l_{0},l_{1},\cdots,l_{h})
} \non \\
&\equiv &
\lim_{n\rightarrow\infty}
\bra 0|(WR_{0})^{l_{h}}
\sigma_{z}^{(i_{h})}
\cdots
(WR_{0})^{l_{1}}
\sigma_{z}^{(i_{1})}
(WR_{0})^{l_{0}}W|0\ket \non \\
&=&
(-1)^{\sum_{s=0}^{h}\lfloor l_{s}/2\rfloor 
+\sum_{s=1}^{h}\alpha_{s}}
{\sin\brace{\cos}}_{\alpha_{0}}((l_{0}+1)\theta)
{\cos\brace{\sin}}_{\alpha_{1}}(l_{1}\theta)
\cdots
{\cos\brace{\sin}}_{\alpha_{h}}(l_{h}\theta) \non \\
&&
\quad\quad
\mbox{for $h=1,2,\cdots$},
\lab{original-asymptotic-rule}
\eeqa
where
$l_{0}=0,1,\cdots$,
$l_{s}=1,2,\cdots$ for $s=1,\cdots,h$,
$\{i_{s}:1\leq i_{s}\leq n$ for $s=1,\cdots,h\}$
are different from each other,
$\lfloor x \rfloor$ represents the largest integer that is less than or equal to $x$,
\beq
\alpha_{s}=l_{s}
(\mbox{mod $2$})
\in\{0,1\}
\quad\quad
\mbox{for $s=0,\cdots,h$},
\eeq
and
\beq
{f\brace{g}}_{\alpha}(x)
\equiv
\left\{
\begin{array}{ll}
f(x) & \mbox{for $\alpha=0$} \\
g(x) & \mbox{for $\alpha=1$}
\end{array}
\right..
\eeq

We show that this holds for $h=1$
in Appendix~\ref{Appendix-subsec-formulas-0-WR0-l-SigmaZ-WR0-k-W0}.
We prove Eq.~(\ref{original-asymptotic-rule})
for arbitrary $h$
by the inductive method.
Let us assume Eq.~(\ref{original-asymptotic-rule})
is satisfied for $h$.
We examine whether it is satisfied for $(h+1)$ or not.

First, we assume $l_{0}=2m$ and $l_{1}=2\tilde{m}$
(both of them are even).
Using Eq.~(\ref{WR0-2l-sigmaZ-WR0-2k-W0}),
we obtain
\beqa
\lefteqn{
{\cal G}^{(h+1)}(2m,2\tilde{m},l_{2},\cdots,l_{h+1})
} \non \\
&=&
\bra 0|(WR_{0})^{l_{h+1}}
\sigma_{z}^{(i_{h+1})}
\cdots
(WR_{0})^{l_{2}}
\sigma_{z}^{(i_{2})}
\biggl[
(WR_{0})^{2\tilde{m}}
\sigma_{z}^{(i_{1})}
(WR_{0})^{2m}W|0\ket
\biggr] \non \\
&\sim &
{\cal G}^{(h)}(2(m+\tilde{m}),l_{2},\cdots,l_{h+1}) \non \\
&&
-\sqrt{2}(-1)^{m}\cos(2m+1)\theta
[-\sqrt{2}\sum_{k=0}^{\tilde{m}-1}
{\cal G}^{(h)}(2k,l_{2},\cdots,l_{h+1}) \non \\
&&
+
\bra 0|(WR_{0})^{l_{h+1}}
\sigma_{z}^{(i_{h+1})}
\cdots
(WR_{0})^{l_{2}}
\sigma_{z}^{(i_{2})}
|\eta_{i_{1}}\ket] \non \\
&&
\quad\quad
\mbox{for $m=0,1,\cdots$, $\tilde{m}=1,2,\cdots$,
as $n\rightarrow \infty$},
\lab{0-WR0-l(k+1)-sigmaZ-even-even-expand}
\eeqa
where we use
$\lim_{n\rightarrow \infty}\cos\theta=1$.
(We eliminate one $\sigma_{z}$ operator
from the above equation.
This technique is used in
Sections~\ref{Section-physical-interpretation-multi-particle-creation}
and
\ref{Section-Matrix-elements-density-operator-second}.)
From now on,
we use a symbol of `$\sim$' as an asymptotic equal sign under
$n\rightarrow \infty$ for a while.
On the other hand,
\beq
\bra 0|(WR_{0})^{l_{h+1}}
\sigma_{z}^{(i_{h+1})}
\cdots
(WR_{0})^{l_{2}}
\sigma_{z}^{(i_{2})}
|\eta_{i_{1}}\ket
=0,
\lab{0-WR0-l(k+1)-sigmaZ-sigmaZ-eta}
\eeq
which is proved in
Appendix~\ref{Appendix-formulas-of-0-WR0-lk-WR0-l1-sigmaZ-eta-i}.
Substituting Eq.~(\ref{original-asymptotic-rule}) for $h$
and Eq.~(\ref{0-WR0-l(k+1)-sigmaZ-sigmaZ-eta})
into Eq.~(\ref{0-WR0-l(k+1)-sigmaZ-even-even-expand}),
and using Eq.~(\ref{Appendix-sum-sin-2l+1-theta}),
we obtain
\beqa
\lefteqn{
{\cal G}^{(h+1)}(2m,2\tilde{m},l_{2},\cdots,l_{h+1})
} \non \\
&\sim &
(-1)^{m+\tilde{m}}
\sin(2m+2\tilde{m}+1)\theta
{\cal F}(l_{2},\cdots,l_{h+1}) \non \\
&&
+2(-1)^{m}\cos(2m+1)\theta
\sum_{k=0}^{\tilde{m}-1}
(-1)^{k}\sin(2k+1)\theta
{\cal F}(l_{2},\cdots,l_{h+1}) \non \\
&=&
(-1)^{m}
[(-1)^{\tilde{m}}
\sin(2m+2\tilde{m}+1)\theta
+2\cos(2m+1)\theta
\frac{(-1)^{\tilde{m}-1}}{2}
\frac{\sin 2\tilde{m}\theta}{\cos\theta}]
{\cal F}(l_{2},\cdots,l_{h+1}) \non \\
&\sim &
(-1)^{m+\tilde{m}}
\sin (2\tilde{m}+1)\theta
\cos 2\tilde{m}\theta
{\cal F}(l_{2},\cdots,l_{h+1})
\quad\quad
\mbox{as $n\rightarrow \infty$}, 
\eeqa
where
\beq
{\cal F}(l_{2},\cdots,l_{h+1})
=
(-1)^{\sum_{s=2}^{h+1}\lfloor l_{s}/2\rfloor 
+\sum_{s=2}^{h+1}\alpha_{s}}
{\cos\brace{\sin}}_{\alpha_{2}}(l_{2}\theta)
\cdots
{\cos\brace{\sin}}_{\alpha_{h+1}}(l_{h+1}\theta).
\lab{definition-asymptotic-function-F}
\eeq
This means Eq.~(\ref{original-asymptotic-rule})
is satisfied for $(h+1)$ with
$(l_{0},l_{1})=(2m,2\tilde{m})$.

Next,
we assume 
$l_{0}=2m$ and $l_{1}=2\tilde{m}+1$
(even and odd).
From Eqs.~(\ref{1-phase-error-WR0-2k-W0})
and
(\ref{WR0-2k+1-eta-i}),
we obtain
\beqa
\lefteqn{
{\cal G}^{(h+1)}(2m,2\tilde{m}+1,l_{2},\cdots,l_{h+1})
} \non \\
&=&
\bra 0|(WR_{0})^{l_{h+1}}
\sigma_{z}^{(i_{h+1})}
\cdots
(WR_{0})^{l_{2}}
\sigma_{z}^{(i_{2})}
\biggl[
(WR_{0})^{2\tilde{m}+1}
\sigma_{z}^{(i_{1})}
(WR_{0})^{2m}W|0\ket
\biggr] \non \\
&\sim &
{\cal G}^{(h)}((2m+2\tilde{m})+1,l_{2},\cdots,l_{h+1}) \non \\
&&
-\sqrt{2}(-1)^{m}\cos(2m+1)\theta
[-\sqrt{2}\sum_{k=0}^{\tilde{m}-1}
{\cal G}^{(h)}(2k+1,l_{2},\cdots,l_{h+1}) \non \\
&&
+\frac{1}{\sqrt{2}}
\bra 0|(WR_{0})^{l_{h+1}}
\sigma_{z}^{(i_{h+1})}
\cdots
(WR_{0})^{l_{2}}
\sigma_{z}^{(i_{2})}
(|0\ket-|\overline{i_{1}}\ket)] \non \\
&&
\quad\quad
\mbox{for $m=0,1,\cdots$, $\tilde{m}=1,2,\cdots$,
as $n\rightarrow \infty$}.
\lab{0-WR0-l(k+1)-sigmaZ-odd-even-expand}
\eeqa
(We eliminate one $\sigma_{z}$ operator
from the above expression
as well as Eq.~(\ref{0-WR0-l(k+1)-sigmaZ-even-even-expand}).)
On the other hand,
\beq
\bra 0|(WR_{0})^{l_{h+1}}
\sigma_{z}^{(i_{h+1})}
\cdots
(WR_{0})^{l_{2}}
\sigma_{z}^{(i_{2})}
(|0\ket-|\overline{i_{1}}\ket)
\sim
{\cal F}(l_{2},\cdots,l_{h+1})
\quad\quad
\mbox{as $n\rightarrow \infty$},
\lab{0-WR0-l(k+1)-sigmaZ-sigmaZ-(0-i1)}
\eeq
which is proved in
Appendix~\ref{Appendix-formulas-of-0-WR0-sigmaZ-sigmaZ-(0-i0)}.
Hence,
substituting
Eq.~(\ref{original-asymptotic-rule}) for $h$
and Eq.~(\ref{0-WR0-l(k+1)-sigmaZ-sigmaZ-(0-i1)})
into Eq.~(\ref{0-WR0-l(k+1)-sigmaZ-odd-even-expand}),
and
using Eq.(\ref{Appendix-sum-cos-2(l+1)-theta}), we obtain
\beqa
\lefteqn{
{\cal G}^{(h+1)}(2m,2\tilde{m}+1,l_{2},\cdots,l_{h+1})
} \non \\
&\sim &
\{(-1)^{m+\tilde{m}}
\cos 2(m+\tilde{m}+1)\theta \non \\
&&
+2(-1)^{m}\cos(2m+1)\theta
[\sum_{k=0}^{\tilde{m}-1}
(-1)^{k}\cos 2(k+1)\theta
-\frac{1}{2}]\}
{\cal F}(l_{2},\cdots,l_{h+1}) \non \\
&=&
(-1)^{m}
\{(-1)^{\tilde{m}}
\cos 2(m+\tilde{m}+1)\theta \non \\
&&
+2\cos(2m+1)\theta
[\frac{1}{2}
+\frac{(-1)^{\tilde{m}-1}}{2\cos\theta}
\cos (2\tilde{m}+1)\theta
-\frac{1}{2}]\}
{\cal F}(l_{2},\cdots,l_{h+1}) \non \\
&\sim &
(-1)^{m+\tilde{m}+1}
\sin (2m+1)\theta
\sin (2\tilde{m}+1)\theta
{\cal F}(l_{2},\cdots,l_{h+1})
\quad\quad
\mbox{as $n\rightarrow \infty$}.
\eeqa
This means Eq.~(\ref{original-asymptotic-rule})
holds for $(h+1)$
with
$(l_{0},l_{1})=(2m,2\tilde{m}+1)$.

In the case of
$(l_{0},l_{1})=(2m+1,2\tilde{m})$, and $(2m+1,2\tilde{m}+1)$,
we can show Eq.~(\ref{original-asymptotic-rule})
is satisfied for $(h+1)$
in similar ways.
Therefore, we obtain Eq.~(\ref{original-asymptotic-rule})
for $h=1,2,\cdots$ by induction.
We pay attention that it does not depend on
$i_{1},\cdots,i_{h}$.

To obtain an asymptotic form of
$\bra 0|T_{h}^{(M)}|0\ket/(Mn)^{h}$,
we take intervals of Figure~\ref{error-diagram}
as $l_{0},\cdots,l_{h-1},2M-\sum_{s=1}^{h-1}l_{s}$,
multiply a factor $n!/(n-h)!$ to the terms
for permutation of $\sigma_{z}$ errors,
and sum up them by $l_{0},\cdots,l_{h-1}$,
\beqa
\lim_{n\rightarrow\infty}
\frac{\bra 0|T_{h}^{(M)}|0\ket}{(Mn)^{h}}
&=&
\lim_{n\rightarrow\infty}
\frac{1}{(Mn)^{h}}
\sum_{l_{0}=0}^{M-h}
\sum_{l_{1}=1}^{M-h-l_{0}}
\cdots
\sum_{l_{h-1}=1}^{M-h-(l_{0}+\cdots+l_{h-2})} \non \\
&&
\times
\frac{n!}{(n-h)!}
|{\cal G}^{(h)}(l_{0},\cdots,l_{h+1},2M-\sum_{s=1}^{h-1}l_{s})|^{2}.
\lab{asymptotic-summation-TkM}
\eeqa
Then, substituting Eq.~(\ref{original-asymptotic-rule}),
$\Theta=\lim_{n\rightarrow\infty}M\theta$,
$\phi_{s+1}=l_{s}\theta$ for $s=0,\cdots,h-1$,
\beq
\lim_{n\rightarrow\infty}
\sum_{l_{0}=0}^{M-h}\theta=\int_{0}^{\Theta}d\phi_{1},
\eeq
and
\beq
\lim_{n\rightarrow\infty}
\sum_{l_{s}=1}^{M-h-\sum_{t=0}^{s-1}l_{t}}\theta
=\int_{0}^{\Theta-\sum_{t=1}^{s}\phi_{t}}d\phi_{s+1}
\quad\quad
\mbox{for $s=1,\cdots,h-1$},
\eeq
into Eq.~(\ref{asymptotic-summation-TkM}),
we obtain
Eqs.~(\ref{Diagrammatic-rule}) and (\ref{Diagrammatic-rule-integral}).

\subsection{Formulas of
$\bra 0|(WR_{0})^{l_{h}}\sigma_{z}^{(i_{h})}
\cdots(WR_{0})^{l_{1}}\sigma_{z}^{(i_{1})}|\eta_{i_{0}}\ket$}
\lab{Appendix-formulas-of-0-WR0-lk-WR0-l1-sigmaZ-eta-i}
In this section,
we show
\beq
\bra 0|(WR_{0})^{l_{h}}\sigma_{z}^{(i_{h})}
\cdots
(WR_{0})^{l_{1}}\sigma_{z}^{(i_{1})}
|\eta_{i_{0}}\ket
=0,
\lab{general-formula-WR0-sigmaZ-eta-i}
\eeq
for $h=1,2,\cdots$,
where
$l_{s}=1,2,\cdots$ for $s=1,\cdots,h$,
and
$\{i_{s}:1\leq i_{s}\leq n
\quad
\mbox{for $s=0,1,\cdots,h$}\}$
are different from each other.
For $h=1$,
it is proved in
Appendix~\ref{Appendix-subsec-formulas-0-WR0-k-SigmaZ-eta}.

To prove Eq.~(\ref{general-formula-WR0-sigmaZ-eta-i})
for arbitrary $h$,
we need to show
\beqa
\lefteqn{
\bra 0|(WR_{0})^{l_{h}}\sigma_{z}^{(i_{h})}
\cdots
(WR_{0})^{l_{1}}\sigma_{z}^{(i_{1})}
|\eta_{j_{1},\cdots,j_{s}}\ket
=0
} \non \\
&&\quad\quad
\mbox{for
$\{i_{1},\cdots,i_{h},j_{1},\cdots,j_{s}\}$
that are different from each other},
\lab{general-formula-WR0-sigmaZ-eta-j1-js} \\
\lefteqn{
\bra 0|(WR_{0})^{l_{h}}\sigma_{z}^{(i_{h})}
\cdots
(WR_{0})^{l_{1}}\sigma_{z}^{(i_{1})}
(|\overline{j_{0},j_{1},\cdots,j_{s}}\ket
-|\overline{j_{1},\cdots,j_{s}}\ket)
=0
} \non \\
&&\quad\quad
\mbox{for
$\{i_{1},\cdots,i_{h},j_{0},j_{1},\cdots,j_{s}\}$
that are different from each other},
\lab{general-formula-WR0-sigmaZ-(j0j1js-j1js)}
\eeqa
where
$1\leq j_{0}\leq n$,
$\cdots$,
$1\leq j_{s}\leq n$,
\beq
|\eta_{j_{1},\cdots,j_{s}}\ket
=\frac{1}{\sqrt{2^{n-s}}}
\sum_{x_{j_{1}}=\cdots=x_{j_{s}}=1}|x\ket,
\eeq
and
\beq
|\overline{j_{0},j_{1},\cdots,j_{s}}\ket
=|0
\cdots 0
\begin{array}[t]{c}
1 \\
\uparrow \\
j_{0}
\end{array}
0
\cdots 0
\begin{array}[t]{c}
1 \\
\uparrow \\
j_{1}
\end{array}
0
\cdots 0
\begin{array}[t]{c}
1 \\
\uparrow \\
j_{s}
\end{array}
0
\cdots 0
\ket.
\eeq
From similar calculations in
Appendix~\ref{Appendix-subsec-formulas-0-WR0-k-SigmaZ-eta},
we can show Eqs.~(\ref{general-formula-WR0-sigmaZ-eta-j1-js})
and (\ref{general-formula-WR0-sigmaZ-(j0j1js-j1js)})
are satisfied for $h=1$.
We prove them for arbitrary $h$ by the inductive method.

Let us derive an explicit form of
\beq
(WR_{0})^{l}\sigma_{z}^{(i)}
|\eta_{j_{1},\cdots,j_{s}}\ket
\quad\quad
\mbox{for $l=1,2,\cdots$},
\eeq
which appears in Eq.~(\ref{general-formula-WR0-sigmaZ-eta-j1-js}).
At first, we consider the case of $s=1$.
We can obtain
\beqa
\sigma_{z}^{(i)}
|\eta_{j}\ket
&=&
\frac{1}{\sqrt{2^{n-1}}}
(\sum_{{y_{j}=1}\atop{y_{i}=0}}
-\sum_{{y_{j}=1}\atop{y_{i}=1}})
|y\ket \non \\
&=&
|\eta_{j}\ket-\sqrt{2}|\eta_{j,i}\ket
\quad\quad
\mbox{for $i\neq j$}.
\lab{sigmaZ-i-eta-j}
\eeqa

Then, we derive an explicit form of
$(WR_{0})^{l}|\eta_{i,j}\ket$.
In the case of $l=1$,
we obtain
\beq
WR_{0}|\eta_{i,j}\ket
=\frac{1}{2}
(|0\ket-|\overline{i}\ket-|\overline{j}\ket+|\overline{i,j}\ket).
\lab{WR0-eta-ij}
\eeq
For $l=2$,
we obtain
\beqa
(WR_{0})^{2}|\eta_{i,j}\ket
&=&
-\frac{1}{2}W
(|0\ket+|\overline{i}\ket+|\overline{j}\ket-|\overline{i,j}\ket) \non \\
&=&
-\frac{1}{2\sqrt{2^{n}}}
(2\sum_{x}|x\ket
-4\sum_{x_{i}=x_{j}=1}|x\ket) \non \\
&=&
-W|0\ket+|\eta_{i,j}\ket.
\eeqa
Hence, we obtain
\beqa
(WR_{0})^{2l}|\eta_{i,j}\ket
&=&
-\sum_{k=0}^{l-1}(WR_{0})^{2k}W|0\ket+|\eta_{i,j}\ket
\quad\quad
\mbox{for $l=0,1,\cdots$},
\lab{WR0-2k-eta-ij} \\
(WR_{0})^{2l+1}|\eta_{i,j}\ket
&=&
-\sum_{k=0}^{l-1}(WR_{0})^{2k+1}W|0\ket
+\frac{1}{2}
(|0\ket-|\overline{i}\ket-|\overline{j}\ket+|\overline{i,j}\ket)
\quad\quad
\mbox{for $l=0,1,\cdots$}.
\lab{WR0-2k+1-eta-ij} 
\eeqa

From Eqs.~(\ref{WR0-2k-eta-i}),
(\ref{sigmaZ-i-eta-j}),
and (\ref{WR0-2k-eta-ij}),
we obtain
\beqa
(WR_{0})^{2l}\sigma_{z}^{(i)}|\eta_{j}\ket
&=&
(WR_{0})^{2l}|\eta_{j}\ket
-\sqrt{2}(WR_{0})^{2l}|\eta_{j,i}\ket \non \\
&=&
-\sqrt{2}\sum_{k=0}^{l-1}(WR_{0})^{2k}W|0\ket+|\eta_{j}\ket
+\sqrt{2}\sum_{k=0}^{l-1}(WR_{0})^{2k}W|0\ket-\sqrt{2}|\eta_{j,i}\ket \non \\
&=&
|\eta_{j}\ket-\sqrt{2}|\eta_{j,i}\ket \non \\
&=&
\sigma_{z}^{(i)}|\eta_{j}\ket
\quad\quad
\mbox{for $l=0,1,\cdots$}.
\lab{WR0-2l-sigmaZ-eta-i}
\eeqa
Furthermore, using Eqs.~(\ref{WR0-2k+1-eta-i}) and (\ref{WR0-2k+1-eta-ij}),
we obtain
\beqa
(WR_{0})^{2l+1}\sigma_{z}^{(i)}|\eta_{j}\ket
&=&
WR_{0}\sigma_{z}^{(i)}|\eta_{j}\ket \non \\
&=&
WR_{0}(|\eta_{j}\ket-\sqrt{2}|\eta_{j,i}\ket) \non \\
&=&
\frac{1}{\sqrt{2}}(|\overline{i}\ket-|\overline{j,i}\ket)
\quad\quad
\mbox{for $l=0,1,\cdots$}.
\eeqa

In general, from similar calculations above,
we can obtain
\beqa
\lefteqn{
(WR_{0})^{2l}\sigma_{z}^{(i)}|\eta_{j_{1},\cdots,j_{s}}\ket
=
|\eta_{j_{1},\cdots,j_{s}}\ket-\sqrt{2}|\eta_{j_{1},\cdots,j_{s},i}\ket
\quad\quad
\mbox{for $l=0,1,\cdots$},
}
\lab{WR0-2l-sigmaZ-eta-j1js} \\
\lefteqn{(WR_{0})^{2l+1}\sigma_{z}^{(i)}|\eta_{j_{1},\cdots,j_{s}}\ket} \\
&=&
\frac{1}{\sqrt{2^{s}}}
\sum_{\alpha\in\{0,1\}^{s}}
(-1)^{\alpha_{1}+\cdots+\alpha_{s}}
|\overline{i,\alpha_{1}j_{1},\cdots,\alpha_{s}j_{s}}\ket \non \\
&=&
\frac{1}{\sqrt{2^{s}}}
\sum_{(\alpha_{2},\cdots,\alpha_{s})\in\{0,1\}^{s-1}}
(-1)^{\alpha_{2}+\cdots+\alpha_{s}}
(|\overline{i,0,\alpha_{2}j_{2},\cdots,\alpha_{s}j_{s}}\ket
-|\overline{i,j_{1},\alpha_{2}j_{2},\cdots,\alpha_{s}j_{s}}\ket)
\non \\
&&
\quad\quad
\mbox{for $l=0,1,\cdots$},
\lab{WR0-(2l+1)-sigmaZ-eta-j1js}
\eeqa
where
\beq
|\overline{i,\alpha_{1}j_{1},\cdots,\alpha_{s}j_{s}}\ket
=|0
\cdots 0
\begin{array}[t]{c}
1 \\
\uparrow \\
i
\end{array}
0
\cdots 0
\begin{array}[t]{c}
\alpha_{1} \\
\uparrow \\
j_{1}
\end{array}
0
\cdots 0
\begin{array}[t]{c}
\alpha_{s} \\
\uparrow \\
j_{s}
\end{array}
0
\cdots 0
\ket.
\eeq
Here, we notice that,
if we relabel indexes,
the right side of Eq.~(\ref{WR0-(2l+1)-sigmaZ-eta-j1js})
is written as a sum of
$(|\overline{j_{0},j_{1},\cdots,j_{r}}\ket
-|\overline{j_{1},\cdots,j_{r}}\ket)$,
which appears in Eq.~(\ref{general-formula-WR0-sigmaZ-(j0j1js-j1js)}).

Then, we derive an explicit form of
\beq
(WR_{0})^{l}\sigma_{z}^{(i)}
(|\overline{j_{0},j_{1},\cdots,j_{s}}\ket
-|\overline{j_{1},\cdots,j_{s}}\ket)
\quad\quad
\mbox{for $l=1,2,\cdots$},
\lab{proposal-WR0-sigmaZ-(j0j1js-j1js)}
\eeq
included by Eq.~(\ref{general-formula-WR0-sigmaZ-(j0j1js-j1js)}).
Here, we can assume
$j_{t}=t$
for $t=1,\cdots,s$,
and
$s<i \leq n$,
without losing generality.
Because
$\{i,j_{0},\cdots,j_{s}\}$
are different from each other,
$\sigma_{z}^{(i)}$ does not have an effect
on the state.
Hence, we obtain
\beqa
\lefteqn{
(WR_{0})^{l}\sigma_{z}^{(i)}
(|\underbrace{1,\cdots,1}_{s+1},\underbrace{0,\cdots,0}_{n-s-1}\ket
-|0,\underbrace{1,\cdots,1}_{s},\underbrace{0,\cdots,0}_{n-s-1}\ket)
} \non \\
&=&
\left\{
\begin{array}{ll}
{\displaystyle
|\underbrace{1,\cdots,1}_{s+1},\underbrace{0,\cdots,0}_{n-s-1}\ket
-|0,\underbrace{1,\cdots,1}_{s},\underbrace{0,\cdots,0}_{n-s-1}\ket
}
&
\mbox{for $l=0,2,\cdots$ (even)} \\
{\displaystyle
\frac{1}{\sqrt{2^{n-2}}}
\sum_{x\in\{0,1\}^{n-1}}
(-1)^{1+x_{1}+\cdots+x_{s}}
|1,x_{1},\cdots,x_{n-1}\ket
}
&
\mbox{for $l=1,3,\cdots$ (odd)}
\end{array}
\right..
\lab{WR0-l-sigmaZ-(110-010)}
\eeqa
From this result,
we find that Eq.~(\ref{proposal-WR0-sigmaZ-(j0j1js-j1js)})
is described as a sum of
$(|\overline{j_{0},j_{1},\cdots,j_{s}}\ket
-|\overline{j_{1},\cdots,j_{s}}\ket)$,
which appears in Eq.~(\ref{general-formula-WR0-sigmaZ-(j0j1js-j1js)}).

Here, we assume Eqs.~(\ref{general-formula-WR0-sigmaZ-eta-j1-js})
and (\ref{general-formula-WR0-sigmaZ-(j0j1js-j1js)})
hold for $h$.
From Eqs.~(\ref{WR0-2l-sigmaZ-eta-j1js}), 
(\ref{WR0-(2l+1)-sigmaZ-eta-j1js}),
and (\ref{WR0-l-sigmaZ-(110-010)}),
we can resolve Eqs.~(\ref{general-formula-WR0-sigmaZ-eta-j1-js})
and (\ref{general-formula-WR0-sigmaZ-(j0j1js-j1js)})
for $(h+1)$
to terms that contain $h$ errors of $\sigma_{z}$,
so that
we can prove them
to be true for $(h+1)$.
Hence,
they are proved for any $h$
by the inductive method.
Therefore, we obtain Eq.~(\ref{general-formula-WR0-sigmaZ-eta-i})
for arbitrary $h$.

\subsection{Formulas of
$\bra 0|(WR_{0})^{l_{h}}\sigma_{z}^{(i_{h})}
\cdots(WR_{0})^{l_{1}}\sigma_{z}^{(i_{1})}
(|0\ket-|\overline{i_{0}}\ket)$}
\lab{Appendix-formulas-of-0-WR0-sigmaZ-sigmaZ-(0-i0)}
In this section,
we show
\beq
\bra 0|(WR_{0})^{l_{h}}\sigma_{z}^{(i_{h})}
\cdots
(WR_{0})^{l_{1}}\sigma_{z}^{(i_{1})}
(|0\ket-|\overline{i_{0}}\ket)
\sim
{\cal F}(l_{1},\cdots,l_{h})
\quad\quad
\mbox{as $n\rightarrow\infty$},
\lab{general-formula-WR0-sigmaZ-(0-i)}
\eeq
for $h=1,2,\cdots$,
where
$l_{s}=1,2,\cdots$ for $s=1,\cdots,h$,
$\{i_{s}:1\leq i_{s}\leq n
\quad
\mbox{for $s=0,1,\cdots,h$}\}$
are different from each other,
and the function ${\cal F}(l_{1},\cdots,l_{h})$
is defined in Eq.~(\ref{definition-asymptotic-function-F})
of Appendix~\ref{Appendix-derivation-asymptotic-forms-matrix-elements}.
Eq.~(\ref{general-formula-WR0-sigmaZ-(0-i)})
is used
for the inductive method
in
Appendix~\ref{Appendix-derivation-asymptotic-forms-matrix-elements}.
It is shown for $h=1$
in Appendix~\ref{Appendix-subsec-formulas-0-WR0-k-SigmaZ-eta}.
We prove it for any $h$ by the inductive method.

We assume Eq.~(\ref{general-formula-WR0-sigmaZ-(0-i)})
is satisfied for $(h-1)$.
Let us consider the following equation,
\beqa
(WR_{0})^{l}\sigma_{z}^{(i)}
(|0\ket-|\overline{j}\ket)
&=&
-(WR_{0})^{l-1}
W(|0\ket+|\overline{j}\ket) \non \\
&=&
-(WR_{0})^{l-1}
(2W|0\ket-\sqrt{2}|\eta_{j}\ket)
\quad\quad
\mbox{for $l=1,2,\cdots$
and $i\neq j$}.
\lab{proposal-WR0-sigmaZ-(0-j)}
\eeqa
Here, we assume $l=2m+1$ (odd).
From Eq.~(\ref{WR0-2k-eta-i}),
we obtain
\beqa
(WR_{0})^{2m+1}\sigma_{z}^{(i)}
(|0\ket-|\overline{j}\ket)
&=&
-2(WR_{0})^{2m}W|0\ket
+\sqrt{2}(WR_{0})^{2m}|\eta_{j}\ket \non \\
&=&
-2(WR_{0})^{2m}W|0\ket
-2\sum_{k=0}^{m-1}(WR_{0})^{2k}W|0\ket
+\sqrt{2}|\eta_{j}\ket \non \\
&=&
-2\sum_{k=0}^{m}(WR_{0})^{2k}W|0\ket
+\sqrt{2}|\eta_{j}\ket.
\eeqa
Using Eq.~(\ref{general-formula-WR0-sigmaZ-eta-i}),
we obtain
\beq
\bra 0|(WR_{0})^{l_{h}}\sigma_{z}^{(i_{h})}
\cdots
(WR_{0})^{l_{2}}\sigma_{z}^{(i_{2})}
(WR_{0})^{2m+1}\sigma_{z}^{(i_{1})}
(|0\ket-|\overline{i_{0}}\ket)
=
-2\sum_{k=0}^{m}
{\cal G}^{(h-1)}(2k,l_{2},\cdots,l_{h}).
\lab{Expansion-0-WR0-sigmaZ-sigmaZ-(0-i0)}
\eeq

Then, we consider the following fact.
We use Eq.~(\ref{general-formula-WR0-sigmaZ-(0-i)})
for obtaining the asymptotic form of
Eq.~(\ref{original-asymptotic-rule})
by induction
in Appendix~\ref{Appendix-derivation-asymptotic-forms-matrix-elements}.
Hence, we can assume that Eq.~(\ref{original-asymptotic-rule})
is true for $(h-1)$.
Therefore, we can require
Eq.~(\ref{Expansion-0-WR0-sigmaZ-sigmaZ-(0-i0)}) as
\beqa
\lefteqn{
\bra 0|(WR_{0})^{l_{h}}\sigma_{z}^{(i_{h})}
\cdots
(WR_{0})^{l_{2}}\sigma_{z}^{(i_{2})}
(WR_{0})^{2m+1}\sigma_{z}^{(i_{1})}
(|0\ket-|\overline{i_{0}}\ket)
} \non \\
&\sim &
-2\sum_{k=0}^{m}
(-1)^{k}\sin(2k+1)\theta{\cal F}(l_{2},\cdots,l_{h}) \non \\
&\sim &
(-1)^{m+1}\sin 2(m+1)\theta{\cal F}(l_{2},\cdots,l_{h}) \non \\
&\sim &
(-1)^{m+1}\sin (2m+1)\theta{\cal F}(l_{2},\cdots,l_{h}) \non \\
&=&
{\cal F}(2m+1,l_{2},\cdots,l_{h})
\quad\quad
\mbox{as $n\rightarrow\infty$},
\eeqa
where we use
Eq.~(\ref{Appendix-sum-sin-2l+1-theta})
and
$\sin 2(m+1)\theta
\sim
\sin (2m+1)\theta$.

Hence, Eq.~(\ref{general-formula-WR0-sigmaZ-(0-i)}) is satisfied
for $h$ in the case that $l$ is odd.
For $l=2m$ (even),
we can give a similar discussion and prove it.
Therefore, we obtain Eq.~(\ref{general-formula-WR0-sigmaZ-(0-i)})
for arbitrary $h$.

\section{Notes for numerical calculations}
\lab{Appendix-notes-for-numerical-calculations}
In this section, we take some notes about numerical calculations
of higher order perturbations.

First, we calculate
the asymptotic forms of the forth and fifth corrections
for density operator.
Using the rules of
Eqs.~(\ref{Diagrammatic-rule})
and (\ref{Diagrammatic-rule-integral}),
we obtain
\beqa
F_{4}(\Theta)
&=&
\frac{1}{48}
+\frac{15-16\Theta^{2}}{12288\Theta^{2}}\cos 4\Theta
-\frac{5(3+32\Theta^{2})}{49152\Theta^{3}}\sin 4\Theta \non \\
&=&
\frac{1}{18}\Theta^{2}+O(\Theta^{3}), \non \\
F_{5}(\Theta)
&=&
\frac{1}{240}
+\frac{45+720\Theta^{2}-256\Theta^{4}}{1966080\Theta^{4}}\cos 4\Theta 
-\frac{3+32\Theta^{2}+256\Theta^{4}}{524288\Theta^{5}}\sin 4\Theta \non \\
&=&
\frac{1}{105}\Theta^{2}+O(\Theta^{3}),
\lab{asymptotic-T45M-integral-form}
\eeqa
where $F_{h}(\Theta)$ is defined in
Eq.~(\ref{Definition-F-h-Theta}).
From the definition of Eq.~(\ref{coefficients-matrix-element-asymptotic-expansion}),
we can obtain $C_{4}(\Theta)$ and $C_{5}(\Theta)$,
the forth and fifth coefficients
of Eq.~(\ref{matrix-element-asymptotic-expansion}).

Next,
we consider the region of $x$ where the perturbation theory
up to the fifth order is valid.
To investigate it,
we examine the sixth order correction,
which is given by
\beqa
F_{6}(\Theta)
&=&
\frac{1}{1440}
+\frac{315+1680\Theta^{2}-256\Theta^{4}}{23592960\Theta^{4}}\cos 4\Theta
-\frac{7(15+256\Theta^{4})}{31457280\Theta^{5}}\sin 4\Theta \non \\
&=&
\frac{\Theta^{2}}{720}+O(\Theta^{3}).
\eeqa
From numerical calculations,
we obtain
\beq
0\leq \frac{1}{6!}C_{6}(\Theta)\leq 1.62\times 10^{-4},
\eeq
where
$0\leq\Theta\leq(\pi/4)$.
Hence, if we limit $x$ to
\beq
0\leq x \leq 1.35,
\lab{reliable-x-region}
\eeq
it is bounded to
\beq
0\leq \frac{1}{6!}C_{6}(\Theta)x^{6}\approx 9.79\times 10^{-4}
\leq 10^{-3}.
\eeq

\end{document}